\documentclass[twocolumn,noshowpacs,amsmath,amssymb]{revtex4-2}
\usepackage{graphicx}
\usepackage{amsfonts}% Include figure files
\usepackage{dcolumn}% Align table columns on decimal point
\usepackage{color}
\usepackage{CJK}
\usepackage{hyperref}

\begin{document}
	
\newcommand{\gin}[1]{{\bf\color{blue}#1}}
\def\bc{\begin{center}}
\def\ec{\end{center}}
\def\bea{\begin{eqnarray}}
\def\eea{\end{eqnarray}}
\newcommand{\avg}[1]{\langle{#1}\rangle}
\newcommand{\Avg}[1]{\left\langle{#1}\right\rangle}
\newcommand{\brm}[1]{\boldsymbol{\mathrm{#1}}}	
	
\title{First passage in discrete-time absorbing Markov chains under stochastic resetting }
\author{Hanshuang Chen$^{1}$}\email{chenhshf@ahu.edu.cn}

\author{Guofeng Li$^{1}$}

\author{Feng Huang{$^{2}$}}

\affiliation{$^{1}$School of Physics and Optoelectronics Engineering, Anhui
	University, Hefei 230601, China \\ $^2$School of Mathematics and Physics, Anhui Jianzhu University, Hefei, 230601, China }

\date{\today}
	
\begin{abstract}	
First passage of stochastic processes under resetting has recently been an active research topic in the field of statistical physics. However, most of previous studies mainly focused on the systems with continuous time and space. In this paper, we study the effect of stochastic resetting on first passage properties of discrete-time absorbing Markov chains, described by a transition matrix $\brm{Q}$ between transient states and a transition matrix $\brm{R}$ from transient states to absorbing states. Using a renewal approach, we exactly derive the unconditional mean first passage time (MFPT) to either of absorbing states, the splitting probability the and conditional MFPT to each absorbing state. All the quantities can be expressed in terms of a deformed fundamental matrix $\brm{Z_{\gamma}}=\left[\brm{I}-(1-\gamma) \brm{Q} \right]^{-1}$ and $\brm{R}$, where $\brm{I}$ is the identity matrix, and $\gamma$ is the resetting probability at each time step. We further show a sufficient condition under which the unconditional MPFT can be optimized by stochastic resetting. Finally, we apply our results to two concrete examples: symmetric random walks on one-dimensional lattices with absorbing boundaries and voter model on complete graphs.   
\end{abstract}
	%\pacs{89.75.Hc, 05.45.-a, 64.60.Cn}
	\maketitle
\section{Introduction}\label{sec1}
First passage underlies a wide variety of stochastic phenomena that have broad applications in phase transitions, neural firing, searching processes, epidemic extinction, and consensus formation, and so on \cite{redner2001guide,van1992stochastic,klafter2011first,bray2013persistence,RevModPhys.85.135}. Recently, first passages under resetting has been an active topic in the field of statistical physics (see \cite{evans2020stochastic} for a recent review), due to its numerous applications spanning across interdisciplinary fields ranging from search problems  \cite{PhysRevLett.113.220602,PhysRevE.92.052127}, the optimization of randomized computer
algorithms \cite{PhysRevLett.88.178701}, and to chemical and biological processes \cite{reuveni2014role,rotbart2015michaelis}. Resetting refers to a sudden interruption of a stochastic process followed by its starting anew.

A canonical diffusion model subject to stochastic
resetting was studied by Evans and Majumdar \cite{evans2011diffusion,evans2011diffusion2}. Resetting can produce a counterintuitive effect: it renders an infinite mean first passage time (MFPT) finite, which can be also minimized at a specific resetting rate. Some extensions have been made in the field, such as temporally or spatially dependent resetting rate \cite{evans2011diffusion2,pal2016diffusion,PhysRevE.96.022130}, higher dimensions \cite{Evans2014_Reset_Highd,arXiv:2109.11101}, complex geometries \cite{Christou2015,PhysRevResearch.2.033027,BressloffJSTAT2021}, noninstantaneous resetting \cite{EvansJPA2018,PalNJP2019,PhysRevE.101.052130,GuptaJPA2020}, in the presence of external potential \cite{pal2015diffusion,ahmad2019first,gupta2020stochastic}, or in the presence of multiple targets \cite{PhysRevE.99.032123,Bressloff2020JPA1,PhysRevE.102.022115}, other types of Brownian motion, like run-to-tumble particles \cite{evans2018run,santra2020run,bressloff2020occupation}, active particles \cite{scacchi2018mean,kumar2020active}, and so on \cite{basu2019symmetric}.  
These nontrivial findings have triggered enormous recent activities in the field, including statistical physics \cite{pal2017first,gupta2014fluctuating,evans2014diffusion,meylahn2015large,chechkin2018random,magoni2020ising}, stochastic thermodynamics \cite{fuchs2016stochastic,pal2017integral,gupta2020work}, and single-particle experiments \cite{tal2020experimental,besga2020optimal}. 

An impressive advantage of resetting is its ability to accelerate the completion of a stochastic process. However, in many cases the resetting can also slow down the completion of a stochastic process. Pal and Reuveni \cite{pal2017first} derived a criterion for restart to be beneficial. They showed restart has the ability to expedite the completion of the underlying stochastic process if the relative standard deviation associated with the first passage time (FPT) without resetting is larger than one. As a corollary, if a non-zero optimal resetting rate exists,
the relative standard deviation is always unity at optimality \cite{PhysRevLett.116.170601}. The resetting criterion was also interpreted by so-called ``inspection paradox'' \cite{Pal2021}. The usefulness of the criterion was demonstrated in systems of a Brownian walker in a one-dimensional domain with and without force field \cite{Ray2019,Durang2019,PhysRevE.99.032123,PhysRevResearch.1.032001,PhysRevE.99.022130,PhysRevE.103.052129}. There has been realizations that so-called ``resetting transition'' occurs at some parameter of the underlying model, which distinguishes that resetting can either hinder or facilitate in the completion of a stochastic process. Resetting transition can be both first \cite{PhysRevLett.113.220602,PhysRevE.92.062115,arXiv:2109.11101} and second order \cite{Christou2015,PhysRevE.97.062106} like in the classical phase transition. A Landau-like theory was also used to characterize phase transitions in resetting systems \cite{PhysRevResearch.1.032001}.

Most of previous works have focused on the systems with continuous time and space. However, the impact of resetting on the systems with discrete time and space has only received less attention. Montero and Villarroel \cite{PhysRevE.94.032132} studies a discrete time unidirectional
random walk on an infinite one-dimensional lattice subject to resetting with a random or site-dependent probability. They analyzed the FPT and survival probabilities for the walker to reach a certain threshold in the lattice. Boyer and Solis-Salas \cite{PhysRevLett.112.240601} proposed a preferential visit model in order to incorporate the memory effect into the resetting processes. The walker either performs a random move locally or relocates to a previously visited site with a probability proportional to the number of past visits to that site. It was shown that the model generates slow sub-diffusion due to the dynamics of memory-driven
resetting. The preferential visit model was further studied in the presence of a single defect site, in which an Anderson-like localization transition was observed \cite{PhysRevLett.119.140603,Boyer2019}.
Majumdar et al.\cite{PhysRevE.92.052126} studied analytically a simple random walk model on a one-dimensional lattice, where at each time step the walker either resets to the maximum of the already visited positions or undergoes symmetric random walks.  They found that for any nonzero resetting probability both the average maximum and the average position grow ballistically with a common velocity. 
Bonomo and Pal \cite{PhysRevE.103.052129} derived a criterion that dictates when restart remains beneficial in discrete space and time restarted processes, and then applied the result to
a symmetric and a biased random walker in one-dimensional lattice confined within two absorbing boundaries.
Riascos et al. \cite{PhysRevE.101.062147} studied random walks on
arbitrary networks subject to resetting with a constant probability.
They derived the exact expressions of the stationary probability
distribution and the MFPT by the spectral representation of
the transition matrix without resetting. Subsequently, the results are
generalized to the case of multiple resetting nodes \cite{PhysRevE.103.062126,Wang2021}. Wald
and B\"ottcher \cite{PhysRevE.103.012122} introduced a framework for studying classical,
quantum, and hybrid random walks with stochastic resetting
on arbitrary networks, in which they derived analytical solutions
of the occupation probability for a classical or quantum
random walker. In a recent work \cite{huang2021random}, we studied random walks on arbitrary networks with first-passage resetting processes \cite{de2020optimization,de2021optimization}, in which we have derived exact expressions of the stationary occupation probability and the MFPT between arbitrary two nonobservable nodes.

In the present work, we aim to study the effect of stochastic resetting on first-passage properties of general absorbing Markovian networks. The underlying network is consisted of absorbing nodes and transient nodes. The system starts from a transient node, and either performs random walks on the Markovian network or is reset to a given transient node with a constant probability. Once the system enters into either of absorbing nodes, the process is terminated. Using a renewal approach, we derive the exact expressions of the unconditional MPFT (uMFPT), splitting probabilities, and conditional MPFT (cMFPT). We also deduce a sufficient condition under which the uMPFT is expedited via stochastic resetting. Finally, we apply our results into two concrete examples: symmetric random walks on one-dimensional lattices with two absorbing endpoints and voter model on complete graphs.

The paper is structured as follows. In Sec.\ref{sec2} and Sec.\ref{sec3}, we present first passage properties of an absorbing Markovian network without and with stochastic resetting, respectively. In Sec.\ref{sec4}, we give a sufficient condition for accelerating uMPFT by stochastic resetting. In Sec.\ref{sec5} we demonstrate our results by two concrete examples. Finally in Sec.\ref{sec6} we provide the conclusions.

\section{Search on an absorbing Markovian network without resetting}\label{sec2}
Let us consider a discrete-time Markovian process between $N$  different states, described by a stochastic matrix $\brm{W}$ whose element $W_{ij}$ gives the transition probability from state $i$ to state $j$. Among $N$ states, there are $m$ ($m < N$) different states that are the searching targets, denoted by $\left\{ {{o_1}, \ldots ,{o_m}} \right\}$. Once the system enters into either of targets, the searching process is terminated. For convenience, the $m$ targets are numbered as the last $m$ states. The transition matrix $\brm{W}$ can be written in the block form,   
\begin{eqnarray}\label{eq1.1}
\brm{W} = \left( {\begin{array}{*{20}{c}}
	\brm{Q}&\brm{R}\\
	\brm{O}&\brm{I}
	\end{array}} \right),
\end{eqnarray}
where $\brm{Q}$ is the $n \times n$ transition matrix between non-target (transient) states (here $n=N-m$ is the number of transient states), and $\brm{R}$ is the $n \times m$ transition matrix from transient states to targets (absorbing states). $\brm{O}$ is the null matrix and $\brm{I}$ is the identity matrix.

Let us denote by $S_i^0(t)$	(here the superscript ``0" denotes the case without resetting) the survival probability of the system up to time $t$ having started from the $i$th transient state, given by  
\begin{eqnarray}\label{eq1.2}
S_i^0\left( t \right) = \sum\limits_{j = 1}^n {{{\left( {{\brm{Q}^t}} \right)}_{ij}}}. 
\end{eqnarray}

We further define $F^0_i(t)$ as the first passage probability that is the probability of the system hits either of targets at time $t$ for the first time. $F^0_i(t)$ can be connected to $S^0_{i}(t)$ by the relation:  
$F^0_{i}(t)=S^0_{i}(t-1)-S^0_{i}(t)$ for $t \ge 1$ and $F^0_{i}(0)=1-S^0_{i}(0)$ for $t =0$.  In the Laplace domain, we have
\begin{eqnarray}\label{eq1.2.1}
{{\tilde F}^0_{i}}(s) =1+ \left( {{e^{ - s}} - 1} \right){{\tilde S}^0_{i}}(s),
\end{eqnarray}
where 
\begin{eqnarray}\label{eq1.2.2}
\tilde S_i^0(s) = \sum\limits_{t = 0}^\infty  {{e^{ - st}}S_i^0\left( t \right)}  = \sum\limits_{j = 1}^n {{{\left[ {{{\left( {\brm{I} - {e^{ - s}}\brm{Q}} \right)}^{ - 1}}} \right]}_{ij}}} .
\end{eqnarray}

The uMFPT from the $i$th transient state to either of targets is given by
\begin{eqnarray}\label{eq1.2.3}
\left\langle {\tau _i^0} \right\rangle  = \sum\limits_{t = 0}^\infty  {tF_i^0\left( t \right) }= - {\left. {\frac{{{d} \tilde F_i^0\left( s \right)}}{{{d} s}}} \right|_{s = 0}} = \tilde S_i^0( 0).
\end{eqnarray}
According to Eq.(\ref{eq1.2.2}), Eq.(\ref{eq1.2.3}) becomes
\begin{eqnarray}\label{eq1.5}
\langle \tau _i^0 \rangle = \sum\limits_{j = 1}^n {{{\left( {{\brm{Z}_0}} \right)}_{ij}}} ,
\end{eqnarray}
where we have defined
\begin{eqnarray}\label{eq1.4}
{\brm{Z}_0} =\brm{ I} + \brm{Q} + {\brm{Q}^2} +  \cdots  = {\left( {\brm{I} - \brm{Q}} \right)^{ - 1}}.
\end{eqnarray}
$\brm{Z}_0$ is called the fundamental matrix associated with $\brm{Q}$. The matrix on the right-hand side of Eq.(\ref{eq1.4}) is called the resolvent of $\brm{Q}$. Eq.(\ref{eq1.5}) states that the uMFPT started from the $i$th transient state is equal to the sum of $i$th row of $\brm{Z}_0$.

Let us denote by $f_{ij}^0(t)$ the probability of the system starting from 
the $i$th transient state and ending at the $j$th absorbing state at time $t$ for the first time. The first-passage probability $f_{ij}^0(t)$ is given by
\begin{eqnarray}\label{eq1.6}
f_{ij}^0\left( t \right) = {\left( {{\brm{Q}^{t - 1}}\brm{R}} \right)_{ij}}.
\end{eqnarray}
Taking the sum for Eq.(\ref{eq1.6}) over $t$, we get the splitting (exit) probability to the $j$th absorbing state having started from the $i$th transient state,
\begin{eqnarray}\label{eq1.7}
\pi _{ij}^0 = \sum\limits_{t = 1}^\infty  {f_{ij}^0\left( t \right)}  = {\left[ {\left( {\brm{I} + \brm{Q} + {\brm{Q}^2} +  \cdots } \right)\brm{R}} \right]_{ij}} = {\left( {{\brm{Z}_0}\brm{R}} \right)_{ij}}. \nonumber \\
\end{eqnarray}
Eq.(\ref{eq1.7}) states that the splitting probability $\pi_{ij}^0$ is equal to the $(i,j)$-entry of the matrix $\brm{Z}_0 \brm{R}$.

The conditional MFPT started from the $i$th transient state and ending in the $j$th absorbing state is given by
\begin{eqnarray}\label{eq1.8}
\langle \tau _{ij}^0 \rangle &=& \frac{1}{{\pi _{ij}^0}}\sum\limits_{t = 1}^\infty  {tf_{ij}^0\left( t \right)} \nonumber \\ &=& \frac{1}{{\pi _{ij}^0}}{\left[ {\left( {\brm{I} + 2\brm{Q} + 3{\brm{Q}^2} +  \cdots } \right)\brm{R}} \right]_{ij}} \nonumber \\ &=& \frac{1}{{\pi _{ij}^0}}{\left( {{\brm{Z}_0^2}\brm{R}} \right)_{ij}}.
\end{eqnarray}

\section{Search on a Markovian network under stochastic resetting}\label{sec3}
We now consider that the system may undergo a resetting process at each time step. With a constant probability $\gamma$, the system is reset to a transient state $r$ (different from any target). With the complementary probability $1-\gamma$, the system goes from one state to another in terms of the transition matrix $\brm{W}$. As long as the system enters into any absorbing state, the process will be terminated.

Let us denote by $S_i(t)$ the survival probability of the system starting from the $i$th transient state until time $t$ in the presence of resetting, which satisfies a first renewal equation \cite{chechkin2018random,pal2016diffusion}, 
\begin{eqnarray}\label{eq2.1}
S_i(t) &=& {\left( {1-\gamma } \right)^t} S_i^0(t) \nonumber \\ &+& \sum_{t'=1}^t {{{\left( {1 - \gamma } \right)}^{t'-1}}\gamma} S_i^0(t'-1){S_r}(t-t') .
\end{eqnarray}
The first term in Eq.(\ref{eq2.1}) corresponds to the case where there is no resetting event at all up to time $t$, which occurs with probability ${\left( {1 - \gamma } \right)^t}$. The second term in Eq.(\ref{eq2.1}) accounts for the event where the first resetting that takes place at time $t'$, which occurs with probability ${\left( {1 - \gamma } \right)^{t'-1}} \gamma$. Before the first resetting, the particle survives with probability $S_{i}^0\left( {t'-1} \right)$, after which the particle survives with probability ${S_{r}}\left( {t - t'} \right)$.

Taking the Laplace transform for Eq.(\ref{eq2.1}), which yields
\begin{eqnarray}\label{eq2.5}
{\tilde S_i}(s) = \tilde S_i^0(s') + \gamma {e^{-s}}\tilde S_i^0(s'){\tilde S_r}(s),
\end{eqnarray}
where $s' = s - \ln \left( {1 - \gamma } \right)$.

Letting $i=r$ in Eq.(\ref{eq2.5}), we have 
\begin{eqnarray}\label{eq2.6}
{{\tilde S}_r}(s) = \frac{{\tilde S_r^0( {s'} )}}{{1 - \gamma {e^{ - s}}\tilde S_r^0( {s'} )}}.
\end{eqnarray}
Substituting Eq.(\ref{eq2.6}) into Eq.(\ref{eq2.5}), we obtain
\begin{eqnarray}\label{eq2.7}
{{\tilde S}_i}( s) = \frac{{\tilde S_i^0( {s'} )}}{{1 - \gamma {e^{ - s}}\tilde S_r^0( {s'})}}.
\end{eqnarray}
Letting $s=0$ in Eq.(\ref{eq2.7}), we obtain the uMFPT in the presence of resetting
\begin{eqnarray}\label{eq2.8}
\langle {\tau _i} \rangle = {{\tilde S}_i}(0) = \frac{{\tilde S_i^0\left( { - \ln \left( {1 - \gamma } \right)} \right)}}{{1 - \gamma \tilde S_r^0\left( { - \ln \left( {1 - \gamma } \right)} \right)}}.
\end{eqnarray}
In Eq.(\ref{eq2.8}), ${\tilde S_i^0\left( { - \ln \left( {1 - \gamma } \right)} \right)}$ can be obtained by Eq.(\ref{eq1.2.2}), 
\begin{eqnarray}\label{eq2.9}
\tilde S_i^0\left( { - \ln \left( {1 - \gamma } \right)} \right) = \sum\limits_{j=1}^{n} { \left(  \brm{Z}_{\gamma} \right)_{ij} } ,
\end{eqnarray}
where we have defined a deformed fundamental matrix
\begin{eqnarray}\label{eq2.10}
{\brm{Z}_\gamma } = {\left[ {\brm{I} - \left( {1 - \gamma } \right)\brm{Q}} \right]^{ - 1}}
\end{eqnarray}
Substituting Eq.(\ref{eq2.9}) into Eq.(\ref{eq2.8}), we obtain, 
\begin{eqnarray}\label{equt}
\langle {\tau _i} \rangle = \frac{{\sum\nolimits_{j = 1}^n {{{\left( {{\brm{Z}_\gamma }} \right)}_{ij}}} }}{{1 - \gamma \sum\nolimits_{j = 1}^n {{{\left( {{\brm{Z}_\gamma }} \right)}_{rj}}} }}.
\end{eqnarray}

Let us denote by $f_{ij}(t)$ the first-passage probability of the system in the presence of resetting, which can establish the connection with $f^0_{ij}(t)$ by the first renewal equation \cite{chechkin2018random,pal2016diffusion}, 
\begin{eqnarray}\label{eq2.11}
{f_{ij}}( t) &= & {\left( {1 - \gamma } \right)^t}f_{ij}^0(t) \nonumber \\ &+& \sum_{t' = 1}^t {{{\left( {1 - \gamma } \right)}^{t' - 1}}\gamma } S_i^0({t'-1}){f_{rj}}( {t - t'} ).
\end{eqnarray}
Taking the Laplace transform for Eq.(\ref{eq2.11}), we obtain
\begin{eqnarray}\label{eq2.15}
{\tilde f_{ij}}(s) = \tilde f_{ij}^0({s'}) + \gamma {e^{ - s}}\tilde S_i^0( {s'}){\tilde f_{rj}}(s),
\end{eqnarray}
where $s' = s - \ln \left( {1 - \gamma } \right)$ as before.

Letting $i=r$ in Eq.(\ref{eq2.15}), we have 
\begin{eqnarray}\label{eq2.16}
{\tilde f_{rj}}(s) = \frac{{\tilde f_{rj}^0(s')}}{{1 - \gamma {e^{ - s}}\tilde S_r^0(s')}}.
\end{eqnarray}
Substituting Eq.(\ref{eq2.16}) into Eq.(\ref{eq2.15}), we obtain
\begin{eqnarray}\label{eq2.17}
{\tilde f_{ij}}(s) = \tilde f_{ij}^0(s') + \frac{{\gamma {e^{ - s}}\tilde S_i^0\left( {s'} \right)}}{{1 - \gamma {e^{ - s}}\tilde S_r^0(s')}}\tilde f_{rj}^0(s').
\end{eqnarray}

The splitting probabilities in the presence of resetting can be deduced by
\begin{eqnarray}\label{eq2.18}
{\pi _{ij}} &=& \sum_{t = 1}^\infty  {{f_{ij}}(t)}  = {{\tilde f}_{ij}}(0) = \tilde f_{ij}^0\left( { - \ln \left( {1 - \gamma } \right)} \right) \nonumber \\ &+& \frac{{\gamma \tilde S_i^0\left( { - \ln ( {1 - \gamma } )} \right)}}{{1 - \gamma \tilde S_r^0\left( { - \ln ( {1 - \gamma } )} \right)}}\tilde f_{rj}^0\left( { - \ln ( {1 - \gamma })} \right) ,
\end{eqnarray}
where $\tilde f_{ij}^0\left( { - \ln ( {1 - \gamma } )} \right)$ can be calculated by Eq.(\ref{eq1.6}), 
\begin{eqnarray}\label{eq2.19}
\tilde f_{ij}^0\left( { - \ln ( {1 - \gamma } )} \right) &=& \left( {1 - \gamma } \right){\left[ {{{\left( {\brm{I} - \left( {1 - \gamma } \right)\brm{Q}} \right)}^{ - 1}}\brm{R}} \right]_{ij}} \nonumber \\ &=& ( {1 - \gamma } ){\left( {{{\brm{Z}}_\gamma }\brm{R}} \right)_{ij}}.
\end{eqnarray}
Substituting Eq.(\ref{eq2.9}) and Eq.(\ref{eq2.19}) into Eq.(\ref{eq2.18}), we obtain the splitting probabilities in the presence of resetting, 
\begin{eqnarray}\label{eq2.192}
\pi _{ij} = ( {1 - \gamma } ){\left( {{\brm{Z}_\gamma }\brm{R}} \right)_{ij}} + \frac{{\gamma \sum\nolimits_{j = 1}^n {{{\left( {{\brm{Z}_\gamma }} \right)}_{ij}}} }}{{1 - \gamma \sum\nolimits_{j = 1}^n {{{\left( {{\brm{Z}_\gamma }} \right)}_{rj}}} }}( {1 - \gamma } ){\left( {{\brm{Z}_\gamma }\brm{R}} \right)_{rj}} . \nonumber \\
\end{eqnarray}

The cMFPT in the presence of resetting is given by
\begin{eqnarray}\label{eq2.20}
\langle {\tau _{ij}} \rangle  = \frac{1}{{{\pi_{ij}}}}\sum_{t = 1}^\infty  {t{f_{ij}}( t)}  =  - \frac{1}{{{\pi _{ij}}}}{\tilde f'_{ij}}(0),
\end{eqnarray}
where $\tilde {f}'_{ij}(0)$ denotes the derivation of $\tilde f_{ij}(s)$ with respect to $s$ at $s=0$,  given by Eq.(\ref{eq2.17})
\begin{widetext}
	\begin{eqnarray}\label{eq2.23}
	{{\tilde f'}_{ij}}(0) &=&  - ( {1 - \gamma }){\left( {\brm{Z}_\gamma ^2\brm{R}} \right)_{ij}} - \frac{{\gamma \sum\nolimits_{j = 1}^n {{{\left( {{\brm{Z}_\gamma }} \right)}_{ij}}} }}{{1 - \gamma \sum\nolimits_{j = 1}^n {{{\left( {{\brm{Z}_\gamma }} \right)}_{rj}}} }}\left( {1 - \gamma } \right){\left( {\brm{Z}_\gamma ^2\brm{R}} \right)_{rj}}  - \gamma ( {1 - \gamma } ){\left( {{\brm{Z}_\gamma }\brm{R}} \right)_{rj}} \nonumber \\ &\times&  \left\{ {\frac{{  ( {1 - \gamma } )\sum\nolimits_{j = 1}^n {{{\left( \brm{Q}  {\brm{Z}_\gamma ^2} \right)}_{ij}}}  + \sum\nolimits_{j = 1}^n {{{\left( {{\brm{Z}_\gamma }} \right)}_{ij}}} }}{{1 - \gamma \sum\nolimits_{j = 1}^n {{{\left( {{\brm{Z}_\gamma }} \right)}_{rj}}} }} + \frac{\gamma {\left[ { ( {1 - \gamma } )\sum\nolimits_{j = 1}^n {{{\left( \brm{Q} {\brm{Z}_\gamma ^2} \right)}_{rj}}}  + \sum\nolimits_{j = 1}^n {{{\left( {{\brm{Z}_\gamma }} \right)}_{rj}}} } \right]{\sum\nolimits_{j = 1}^n {{{\left( {{Z_\gamma }} \right)}_{ij}}}}}}{{{{\left[ {1 - \gamma \sum\nolimits_{j = 1}^n {{{\left( {{\brm{Z}_\gamma }} \right)}_{rj}}} } \right]}^2}}}} \right\} .
	\end{eqnarray}
\end{widetext}
During the derivation of Eq.(\ref{eq2.23}), we have used 
\begin{eqnarray}\label{eq2.21}
\tilde {S'}_i^0\left( { - \ln \left( {1 - \gamma } \right)} \right) =  -\left( {1 - \gamma } \right)\sum\limits_{j=1}^{n} {{{\left( {\brm{Q} \brm{Z}_{\gamma} ^2} \right)}_{ij}}} ,
\end{eqnarray}
and 
\begin{eqnarray}\label{eq2.22}
\tilde {f'}_{ij}^0\left( { - \ln ( {1 - \gamma } )} \right) = - ( {1 - \gamma } ){\left( {\brm{Z}_\gamma ^2 \brm{R}} \right)_{ij}}.
\end{eqnarray}

\section{Condition for optimizing $u$MFPT by stochastic resetting}\label{sec4}
To search the condition for optimizing $\langle \tau_i \rangle$ by stochastic resetting, we take the derivative for Eq.(\ref{equt}) with respect to $\gamma$, which yields
\begin{eqnarray}\label{eq4.1}
\frac{{\partial \langle {{\tau _i}} \rangle }}{{\partial \gamma }} &=& \frac{{\sum\nolimits_{j = 1}^n {{{\left( {{\brm{Z}_\gamma }} \right)}_{ij}}} \left[ {\sum\nolimits_{j = 1}^n {{{\left( {{\brm{Z}_\gamma }} \right)}_{rj}}}  - \gamma \sum\nolimits_{j = 1}^n {{{\left( {\brm{Q} \brm{Z}_\gamma ^2} \right)}_{rj}}} } \right]}}{{{{\left[ {1 - \gamma \sum\nolimits_{j = 1}^n {{{\left( {{\brm{Z}_\gamma }} \right)}_{rj}}} } \right]}^2}}} \nonumber \\ &-& \frac{{\sum\nolimits_{j = 1}^n {{{\left( {\brm{Q}\brm{Z}_\gamma ^2} \right)}_{ij}}} }}{{1 - \gamma \sum\nolimits_{j = 1}^n {{{\left( {{\brm{Z}_\gamma }} \right)}_{rj}}} }},
\end{eqnarray}
where we have used $\partial {\brm{Z}_\gamma }/\partial \gamma  = -\brm{Q}\brm{Z}_\gamma ^2$ in terms of Eq.(\ref{eq2.10}). At $\gamma=0$, Eq.(\ref{eq4.1}) becomes
\begin{eqnarray}\label{eq4.2}
\Delta={\left. {\frac{{\partial \langle {{\tau _i}} \rangle }}{{\partial \gamma }}} \right|_{\gamma  = 0}} &=& {{\sum_{j = 1}^n {{{\left( {{\brm{Z}_0}} \right)}_{ij}}} \sum_{j = 1}^n {{{\left( {{\brm{Z}_0}} \right)}_{rj}}}  - \sum_{j = 1}^n {{{\left( {\brm{Q}\brm{Z}_0 ^2} \right)}_{ij}}} }} \nonumber \\& =& \langle {\tau _i^0} \rangle \langle {\tau _r^0} \rangle- \sum_{j = 1}^n {{{\left( {\brm{Q} \brm{Z}_0^2} \right)}_{ij}}}.
\end{eqnarray}
If $\Delta<0$, $\langle {\tau _i} \rangle $ will decrease when the resetting probability $\gamma$ is increased from zero.  Furthermore, $\langle {\tau _i} \rangle $ will diverge in the limit of $\gamma \to 1$. Assembling these information, one can assert that there exists an optimal resetting probability, $\gamma=\gamma_{\rm{opt}}$, at which $\langle {\tau _i} \rangle $ is a minimum.

However, we should stress that $\Delta>0$ does not necessarily imply that resetting cannot expedite the completion of the searching process. While in this latter case the introduction of a small resetting probability will surely increase
the mean completion time, resetting with an intermediate
probability may still expedite completion \cite{PhysRevLett.113.220602,PhysRevE.92.062115,arXiv:2109.11101}.

\section{Applications}\label{sec5}
\subsection{Symmetric random walks on one-dimensional lattices with absorbing boundaries}

\begin{figure*}
	\centerline{\includegraphics*[width=1.8\columnwidth]{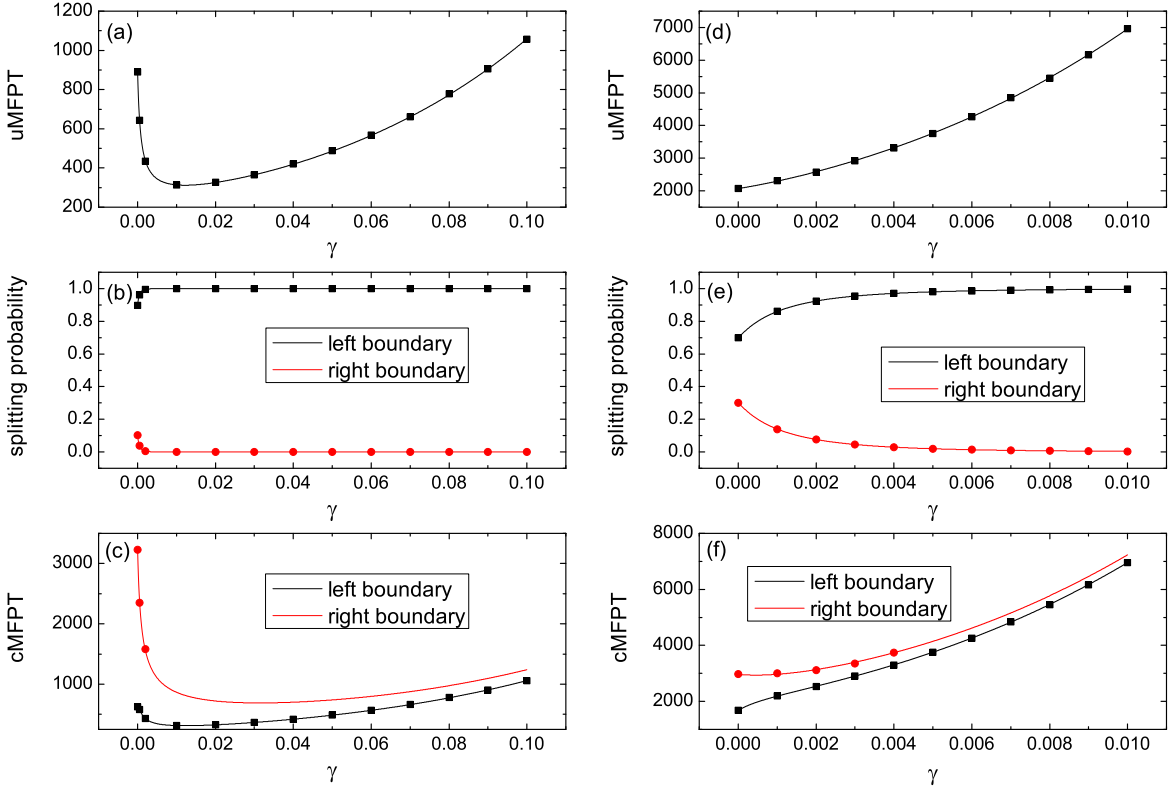}}	\caption{Results of symmetric random walks on a one-dimensional lattice of size $N=100$, where the walker starts from a site $i=10$ (left panel) or from a site $i=30$ (right panel). (a) and (d): The uMFPT as a function of  resetting probability $\gamma$. (b) and (e): the splitting probability to left end or to right end as a function of $\gamma$. (c) and (f): the cMFPT to left end or to right end as a function $\gamma$. Lines and symbols represent the theoretical and simulation results, respectively.  \label{fig0}}
\end{figure*}

We consider a symmetric random walk on a one-dimensional lattice of size $N=n+2$, where both ends are set to be absorbing boundaries. At each time step, the walker hops to either left or to right with equal  probability $\frac{1}{2}$. The $n \times n$ transition matrix $\brm{Q}$ between transient states and the $n \times 2$ transition matrix $\brm{R}$ from transient states to absorbing states can be written as 
\begin{equation}\label{eq5.1}
\brm{Q} = \left( {\begin{array}{*{20}{c}}
	0&{1/2}&{}&{}&{}\\
	{1/2}&0&{1/2}&{}&{}\\
	{}&{1/2}& \ddots & \ddots &{}\\
	{}&{}& \ddots & \ddots &{1/2}\\
	{}&{}&{}&{1/2}&0
	\end{array}} \right),
\end{equation}
and
\begin{equation}\label{eq5.12}
\brm{R} = \left( {\begin{array}{*{20}{c}}
	{1/2}&0\\
	\vdots & \vdots \\
	0&{1/2}
	\end{array}} \right).
\end{equation}
respectively.
We can see that $\brm{Q}$ is a symmetric tridiagonal matrix. $\brm{I}-\brm{Q}$ is also a symmetric tridiagonal matrix, and its inverse $\brm{Z}_0$ can be obtained explicitly \cite{usmani1994inversion},
\begin{equation}\label{eq5.2}
{(\brm{Z}_0)_{ij}} = \frac{{2\min \left\{ {i,j} \right\}\left( {n+1 - \max \left\{ {i,j} \right\}} \right)}}{n+1}.
\end{equation}
Subsituting Eq.(\ref{eq5.2}) into Eq.(\ref{eq1.5}), we obtain
\begin{eqnarray}\label{eq5.3}
\langle \tau _i^0 \rangle = \sum\limits_{j = 1}^n {{{\left( {{\brm{Z}_0}} \right)}_{ij}}}=(n+1-i)i .
\end{eqnarray}

\begin{eqnarray}\label{eq5.4}
\sum\limits_{j = 1}^n {{{\left( {\brm{Z}_0^2} \right)}_{ij}}} =\frac{i}{6}\left[ {2 - i + {i^3} + 4n + 3{n^2} + {n^3} - 2{i^2}\left( {1 + n} \right)} \right]. \nonumber \\
\end{eqnarray}

\begin{eqnarray}\label{eq5.5}
\sum_{j=1}^n {{{\left( {\brm{Q}\brm{Z}_0^2} \right)}_{ij}}}  = \frac{i}{6}\left[ { - 4 + 5i + {i^3} - 2n + 3{n^2} + {n^3} - 2{i^2}( {1 + n} )} \right]. \nonumber \\
\end{eqnarray}

Substituting Eq.(\ref{eq5.3}) and Eq.(\ref{eq5.5}) into Eq.(\ref{eq4.2}), we obtain
\begin{eqnarray}\label{eq5.6}
{\Delta}& =& \frac{i}{6}( {n + 1 - i} ) \nonumber \\ &\times &\left[ 4+i(i-n-1)-n(n+2)+6r(n+1-r) \right]\nonumber \\
\end{eqnarray}
If the resetting node is the same as the original one, $r=i$, 
Eq.(\ref{eq5.6}) leads to the sufficient condition for optimization by stochastic resetting,
\begin{eqnarray}\label{eq5.7}
i<i_{c_1} \quad   {\rm{or}} \quad  i>i_{c_2},
\end{eqnarray}
with
\begin{eqnarray}\label{eq5.71}
{i_{{c_{1(2)}}}} = \frac{{n + 1}}{2} \mp \frac{1}{2}\sqrt {\frac{{{n^2} + 2n + 21}}{5}}.
\end{eqnarray}
In the limit of $n \to \infty$, Eq.(\ref{eq5.71}) simplifies to
\begin{eqnarray}\label{eq5.8}
{i_{{c_{1(2)}}}} = \frac{{5 \mp \sqrt 5 }}{{10}}\left( {n + 1} \right).
\end{eqnarray}
If the length between neighboring site is equal to one, $n+1$ is the total length of the lattice, and then Eq.\ref{eq5.8} recovers to the result on the continuous case \cite{PhysRevE.97.062106,Durang2019,PhysRevE.99.032123}.  

\begin{figure}
	\centerline{\includegraphics*[width=1.0\columnwidth]{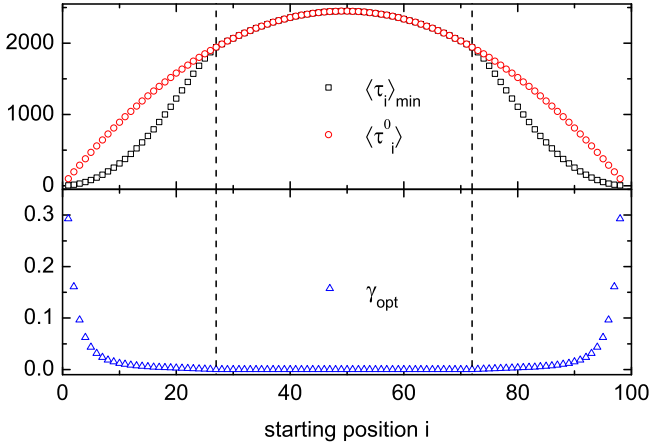}}	\caption{Results on one-dimensional symmetric random walks. (a) The minimum of uMFPT in the presence of resetting, $\langle \tau_i \rangle_{\min}$, and the uMFPT without resetting, $\langle \tau_i^0 \rangle$, as a function of the starting position $i$ of the walker. (b) The optimal resetting probability $\gamma_{\rm{opt}}$ as a function of the starting position $i$ of the walker. The vertical dashed lines indicate the locations of $i_{c_{1(2)}}$.  \label{fig1}}
\end{figure}

When the optimization condition in Eq.(\ref{eq5.7}) holds, there exists an optimal resetting probability, $\gamma=\gamma_{\rm{opt}}$, at which the uMFPT is a minimum. For example, for a one-dimensional lattice with $N=100$ sites, $\langle \tau_i \rangle$ can be optimized by stochastic resetting when $i<28$ or $i>72$. In Fig.\ref{fig0}, we show the uMFPT, splitting probabilities, and cMFPT as a function of resetting probability $\gamma$ for two different starting position: $i=10$ (left panel) and $i=30$ (right panel). Simulation results (symbols) finds excellent agreement with the theoretical ones (lines). In Fig.\ref{fig1}(a), we compare the minimum of the uMFPT in the presence of resetting with the uMFPT without resetting. As expected, when the starting node $i$ of the walker is close to either end the uMFPT can be optimized by resetting and attains a minimum at $\gamma=\gamma_{\rm{opt}}$. Otherwise, the resetting is not beneficial for accelerating the uMFPT. In Fig.\ref{fig1}(b), we show that the value of $\gamma_{\rm{opt}}$ decreases monotonically and becomes zero until the condition in Eq.(\ref{eq5.7}) is no longer satisfied.

For the system, the uMFPT can be also obtained explicitly by calculating the elements of $\brm{Z}_{\gamma}$, given by
\begin{eqnarray}\label{eq5.9}
{\left( {{\brm{Z}_\gamma }} \right)_{ij}} = \left\{ \begin{array}{lr}
\frac{{{{\left( {1 - \gamma } \right)}^{j - i}}{\theta _{i - 1}}{\phi _{j + 1}}}}{{{2^{j - i}}{\theta _n}}}, &i < j\\
\frac{{{\theta _{i - 1}}{\phi _{j + 1}}}}{{{\theta _n}}},&i = j\\\frac{{{{\left( {1 - \gamma } \right)}^{i - j}}{\theta _{j - 1}}{\phi _{i + 1}}}}{{{2^{i - j}}{\theta _n}}},&i > j
\end{array} \right.
\end{eqnarray}
where
\begin{eqnarray}\label{eq5.10}
\left\{ \begin{array}{l}
{\theta _i} = \frac{{{\mu _2} - 1}}{{{\mu _2} - {\mu _1}}}\mu _1^i + \frac{{1 - {\mu _1}}}{{{\mu _2} - {\mu _1}}}\mu _2^i\\
{\phi _i} = \frac{{{\nu _2} - 1}}{{ {{\nu _2} - {\nu _1}} }}\nu _1^{i-n} + \frac{{1 - {\nu _1}}}{{ {{\nu _2} - {\nu _1}} }}\nu _2^{i-n}
\end{array} \right.
\end{eqnarray}
with
\begin{eqnarray}\label{eq5.11}
\left\{ \begin{array}{l}
{\mu _{1,2}} = \frac{1}{2}\left( {1 \pm \sqrt {2\gamma  - {\gamma ^2}} } \right)\\
{\nu _{1,2}} = \frac{2}{{{{\left( {1 - \gamma } \right)}^2}}}\left( {1 \pm \sqrt {2\gamma  - {\gamma ^2}} } \right)
\end{array} \right.
\end{eqnarray}

\subsection{Voter model on complete graphs}
\begin{figure*}
	\centerline{\includegraphics*[width=1.8\columnwidth]{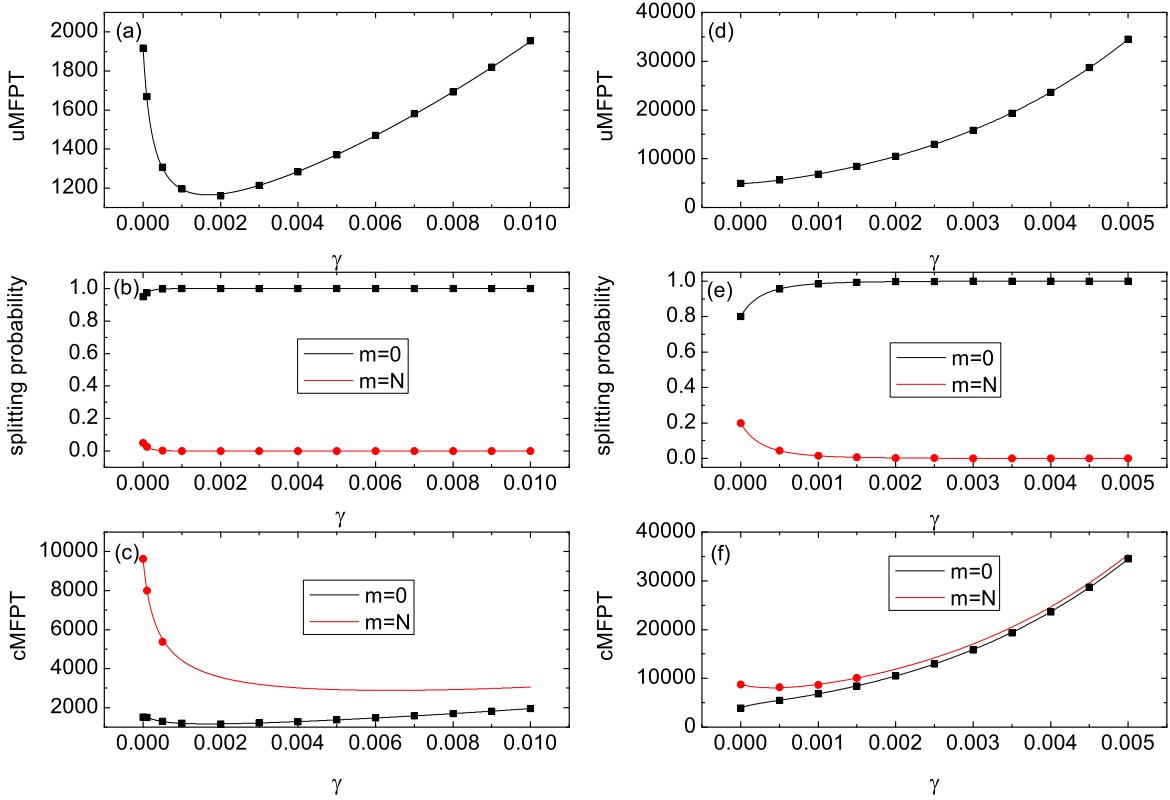}}	\caption{Results of voter model on a complete graph of size $N=100$ with two different initial number $m_0$ of voters with state 1:  $m_0=5$ (left panel) or $m_0=20$ (right panel). (a) and (d): The uMFPT as a function of  resetting probability $\gamma$. (b) and (e): the splitting probability to $m=0$ (all voters with state 0) or to $m=N$ (all voters with state 1) as a function of $\gamma$. (c) and (f): the cMFPT to $m=0$ (all voters with state 0) or to $m=N$ (all voters with state 1) as a function $\gamma$. Lines and symbols respectively represent the theoretical and simulation results, and they are excellent agreement.  \label{fig4}}
\end{figure*}

We consider voter model on a complete graph of size $N$ \cite{PhysRevLett.94.178701,PhysRevE.77.041121}. 
Each node can be in one of two discrete states: 0 and 1. In each time step, a node is randomly chosen, and it adopts the state of a random neighbor. Let us denote $m$ the number of nodes with state 1, with $0 \leq m \leq N$. $m=0$ and $m=N$ are two absorbing states, corresponding to all nodes achieving consensus. The model can be viewed as random walks in the $m$-space. The element of transition matrix is 
\begin{eqnarray}\label{eq6.1}
{W_{m,m'}} &=& \frac{{\left( {N - m} \right)m}}{{N\left( {N - 1} \right)}}{\delta _{m',m + 1}} + \frac{{m\left( {N - m} \right)}}{{\left( {N - 1} \right)N}}{\delta _{m',m - 1}} \nonumber \\ &+& \left[ {1 - 2\frac{{\left( {N - m} \right)m}}{{N\left( {N - 1} \right)}}} \right]{\delta _{m',m}}    .
\end{eqnarray}

From Eq.(\ref{eq6.1}), one can extract the $(N-1) \times (N-1)$ transition matrix $\brm{Q}$ between transient states, given by
\begin{eqnarray}
\brm{Q}=\left( {\begin{array}{*{20}{c}}
	{1 - 2{g_1}}&{{g_1}}&{}&{}&{}\\
	{{g_2}}&{1 - 2{g_2}}&{{g_2}}&{}&{}\\
	{}&{{g_3}}& \ddots & \ddots &{}\\
	{}&{}& \ddots & \ddots &{{g_{N - 2}}}\\
	{}&{}&{}&{{g_{N - 1}}}&{1 - 2{g_{N - 1}}}
	\end{array}} \right)  ,
\end{eqnarray}
and the $(N-1) \times 2$ transition matrix $\brm{R}$ from transient states to absorbing states 
\begin{eqnarray}
\brm{R} = \left( {\begin{array}{*{20}{c}}
	{{g_1}}&0\\
	\vdots & \vdots \\
	0&{{g_{N - 1}}}
	\end{array}} \right).
\end{eqnarray}
where $g_m= \frac{{m\left( {N - m} \right)}}{{N\left( {N - 1} \right)}}$ with $m=1,\cdots,N-1$. 

The model is initialized with $m_0$ voters with state 1, and the remaining voters with state 0. When the system has not reached the consensus state, it evolves in the following way. At each time step, the model is updated according to the usual rule as mentioned before with the probability $1-\gamma$. With the complementary probability $\gamma$, the model is reset to the initial configuration and then the process starts anew. Once the model enters into either of two absorbing states, the process is terminated. 
 
\begin{figure}
	\centerline{\includegraphics*[width=1.0\columnwidth]{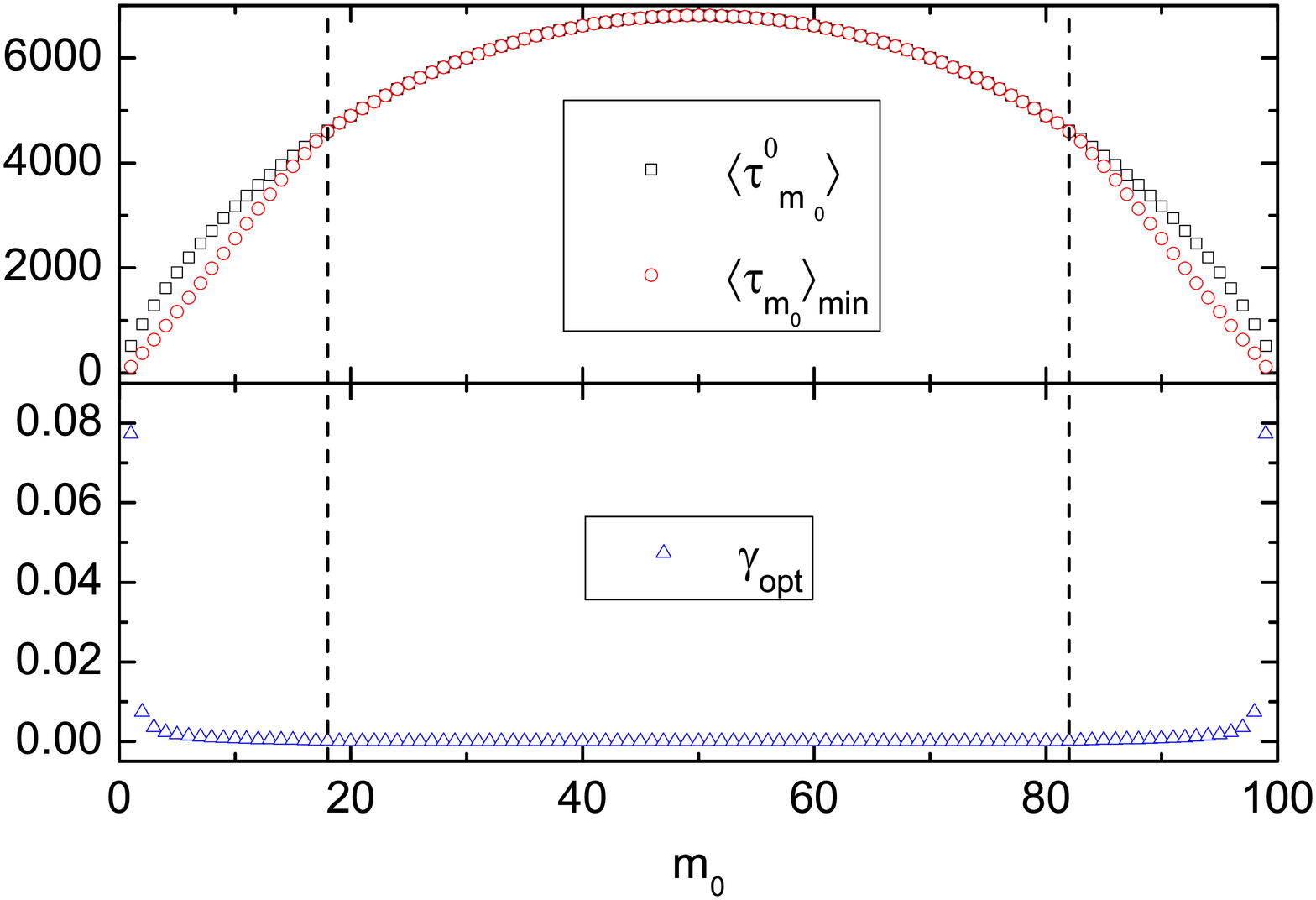}}
	\caption{Results of voter model on a complete graph of size $N=100$. (a) The minimum of uMFPT in the presence of resetting, $\langle \tau_i \rangle_{\min}$, and the conditional MFPT without resetting, $\langle \tau_i^0 \rangle$, as a function of the initial number $m_0$ of voters with state 1. (b) The optimal resetting probability $\gamma_{\rm{opt}}$ as a function of $m_0$. The vertical dashed lines indicate the locations of $m_{c_{1(2)}}$.  \label{fig2}}
\end{figure}

In Fig.\ref{fig4}, we show the uMFPT, splitting probabilities and cMFPT to two absorbing states as a function of resetting probability $\gamma$ for two different initial conditions on a complete graph with $N=100$ nodes. When the initial number $m_0$ of voters with state 1 is close to zero or the total number $N$ of voters, the uMFPT attains a minimum at $\gamma=\gamma_{\rm{opt}}$ (see left panel of Fig.\ref{fig4}), implying that the resetting can optimize the uMFPT. Otherwise, the uMFPT shows a monotonic increase with $\gamma$, and thus the resetting is against the acceleration of the uMFPT (see right panel of Fig.\ref{fig4}). By Eq.(\ref{eq4.2}), we find that when $m_0 \leq m_{c_1}=18 $ or $m_0 \geq m_{c_2}=82$, the uMFPT, i.e. the mean time to consensus, can be optimized by the stochastic resetting. That is to say, when $m_0 \leq m_{c_1} $ or $m_0 \geq m_{c_2}$, there exists a nonzero resetting probability $\gamma_{\rm{opt}}$ at which the uMFPT is a minimum. The is clearly realized from Fig.\ref{fig2}(a), in which we show the minimum of the uMFPT in the presence of resetting, $\langle \tau_{m_0} \rangle_{\min}$, and the uMFPT $\langle \tau_{m_0}^0 \rangle$ in the resetting-free process, as a function of $m_0$. $\gamma_{\rm{opt}}$ decreases as $m_0$ approaches $m_{c_1}$ from below or $m_{c_2}$ from above, as shown in Fig.\ref{fig2}(b). In the limit of $N \to \infty$, the voter model can be described a diffusion-like equation, from which we can solve the survival probability that the system has not reached
the fully ordered state up to time $t$, and then the uMFPT and the unconditional mean squared FPT. According to the condition for optimization by resetting in the continuous version \cite{pal2017first}, we can obtain the values of $m_{c_{1(2)}}/N$ in the limit of $N \to \infty$ (see Appendix \ref{app1} for details),
\begin{eqnarray}\label{eq48}
m_{c_1}/N=1-m_{c_2}/N=0.1846905.
\end{eqnarray}

\begin{figure}
	\centerline{\includegraphics*[width=1.0\columnwidth]{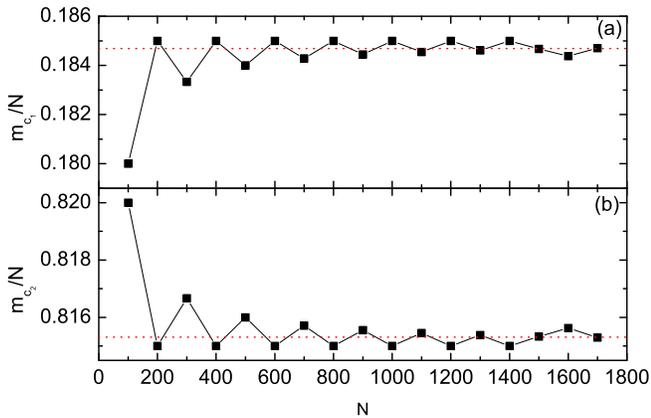}}
	\caption{$m_{c_{1(2)}}/N$ as a function of network size $N$. The horizontal dotted lines indicate the values of $m_{c_{1(2)}}/N$ in the limit of $N \to \infty$ given in Eq.(\ref{eq48}).   \label{fig3}}
\end{figure}

In Fig.\ref{fig3}, we show the values of $m_{c_{1(2)}}/N$ as a function of $N$, from which we can see that $m_{c_{1(2)}}/N$ converge to the limiting values given in Eq.(\ref{eq48}) in an oscillatory way.  

\section{Conclusions}\label{sec6}
In conclusion, we have studied the impact of stochastic resetting on the first passage of a general absorbing Markovian network. Thanks to the renewal structures, we have established the connection between first passage properties with and without resetting. Based on the connection, we have derived exact expressions of the uMPFT, splitting probabilities, and cMPFT as a function of resetting probability. Furthermore, we present a sufficient condition under which the resetting can expedite the uMFPT. Finally, we apply our results to two typical examples: symmetric random walks on one-dimensional lattices with two absorbing ends, and the voter model on complete graphs. In the two examples, we have found that when the initial condition is prepared such that it is close to either of absorbing states, the resetting is able to accelerate the completion of the underlying stochastic process. In the two examples, we have also shown explicitly the conditions for acceleration via resetting. When the size of system tends to infinity, the conditions recover to their counterparts in the continuous case. In the future, we hope that our results can be applied to more complex situations, which may inspire practical implications in stochastic processes by taking advantage of restart.

\begin{acknowledgments}
	This work is supported by the National Natural Science Foundation of China (Grants No. 11875069, No 61973001) and the Key Scientific Research Fund of Anhui Provincial Education Department under (Grant No. KJ2019A0781).
\end{acknowledgments}

\appendix
\section{Derivation of optimization condition for the voter model in the continous limit}\label{app1}
In the continuous limit ($N \to \infty$), the voter model can be described by the Fokker-Planck equation \cite{PhysRevLett.94.178701,PhysRevE.77.041121},
\begin{eqnarray}\label{eq6.2}
\frac{{\partial c\left( {x,t} \right)}}{{\partial t}} = \frac{1}{N}\frac{{{\partial ^2}}}{{\partial {x^2}}}\left[ {\left( {1 - {x^2}} \right)c\left( {x,t} \right)} \right],
\end{eqnarray}
where $x=2m/N-1 \in [-1,1]$ and $c(x,t)$ is the probability density of the system having $x$ at time $t$. The general solution of Eq.(\ref{eq6.2}) is given by the series expansion \cite{AvrahamJPA1990,SlaninaEPJB2003,VazquezNJP2008}
\begin{eqnarray}\label{eq6.3}
c( {x,t} ) = \sum\limits_{l = 0}^\infty  {{A_l}C_l^{3/2}(x){e^{ - \frac{{\left( {l + 1} \right)\left( {l + 2} \right)}}{N}t}}} ,
\end{eqnarray}
where $A_l$ are coefficients determined by the initial condition and $C_l^{3/2}(x)$ are the Gegenbauer polynomials. For the initial condition $c(x,0)=\delta(x-x_0)$, the coefficients $A_l$ are given by
\begin{eqnarray}\label{eq6.4}
{A_l} = \frac{{( {2l + 3} )(1-x_0^2)C_l^{3/2}(x_0)}}{{2(l+1)(l+2)}}.
\end{eqnarray}

The survival probability is given by
\begin{eqnarray}\label{eq6.5}
{S^0}\left( {t|{x_0}} \right) = \int_{ - 1}^1 {c\left( {x,t} \right)dx = } \sum\limits_{l = 0}^\infty  {{A_l}{B_l}{e^{ - \frac{{\left( {l + 1} \right)\left( {l + 2} \right)}}{N}t}}} ,
\end{eqnarray}
where
\begin{eqnarray}\label{eq6.6}
{B_l} = \int_{ - 1}^1 {C_l^{3/2}\left( x \right)dx = } 1 - {\left( { - 1} \right)^{l + 1}}.
\end{eqnarray}

The Laplace transform of ${S^0}(t|x_0 )$ is 
\begin{eqnarray}\label{eq6.7}
{{\tilde S}^0}(s|x_0) = \sum_{l = 0}^\infty  {\frac{{N{A_l}{B_l}}}{{2 + l( {l + 3} ) + Ns}}}.
\end{eqnarray}
The MFPT is
\begin{eqnarray}\label{eq6.8}
\langle {{\tau ^0}( {{x_0}} )} \rangle  = {{\tilde S}^0}( 0 ) = \sum_{l = 0}^\infty  {\frac{{N{A_l}{B_l}}}{{2 + l( {l + 3} )}}} ,
\end{eqnarray}
and 
the mean squared FPT is 
\begin{eqnarray}\label{eq6.9}
\langle {{{\left( {{\tau ^0}\left( {{x_0}} \right)} \right)}^2}} \rangle  =  - 2{ {\frac{{d{{\tilde S}^0}( s )}}{{ds}}} |_{s = 0}} = \sum_{l = 0}^\infty  {\frac{{2{N^2}{A_l}{B_l}}}{{{{\left[ {2 + l( {l + 3} )} \right]}^2}}}}. 
\end{eqnarray}

If we only consider $l=0$ (corresponding to the slowest eigenmode), we find that the condition for optimization
becomes \cite{pal2017first}
\begin{eqnarray}\label{eq6.10}
{\Delta } &=& 2 \langle {{\tau ^0}( {{x_0}} )} \rangle^2- \langle {{{\left( {{\tau ^0}\left( {{x_0}} \right)} \right)}^2}} \rangle \nonumber \\ &=& \frac{3}{8}\left( {1 - 4x_0^2 + 3x_0^4} \right) < 0,
\end{eqnarray}
leading to 
\begin{eqnarray}\label{eq5.20}
x_0\in (-1,-0.57735)  \cup (0.57735, 1),
\end{eqnarray}
or equivalently, 
\begin{eqnarray}\label{eq5.21}
m_0/N \in (0,0.211325)  \cup (0.788675, 1).
\end{eqnarray}

Furthermore, we consider much more eigenmodes, $l=0,2,\cdots,100$ (in terms of Eq.(\ref{eq6.6}) only the even values of $l$ contribute to the sum in Eq.(\ref{eq6.7})), we obtain
\begin{eqnarray}\label{eq5.22}
x_0\in (-1,-0.630619)  \cup (0.630619, 1)
\end{eqnarray}
or equivalently, 
\begin{eqnarray}\label{eq5.23}
m_0/N \in (0,0.1846905)  \cup (0.8153095, 1)
\end{eqnarray}


\begin{thebibliography}{74}
	\expandafter\ifx\csname natexlab\endcsname\relax\def\natexlab#1{#1}\fi
	\expandafter\ifx\csname bibnamefont\endcsname\relax
	\def\bibnamefont#1{#1}\fi
	\expandafter\ifx\csname bibfnamefont\endcsname\relax
	\def\bibfnamefont#1{#1}\fi
	\expandafter\ifx\csname citenamefont\endcsname\relax
	\def\citenamefont#1{#1}\fi
	\expandafter\ifx\csname url\endcsname\relax
	\def\url#1{\texttt{#1}}\fi
	\expandafter\ifx\csname urlprefix\endcsname\relax\def\urlprefix{URL }\fi
	\providecommand{\bibinfo}[2]{#2}
	\providecommand{\eprint}[2][]{\url{#2}}
	
	\bibitem[{\citenamefont{Redner}(2001)}]{redner2001guide}
	\bibinfo{author}{\bibfnamefont{S.}~\bibnamefont{Redner}},
	\emph{\bibinfo{title}{A guide to first-passage processes}}
	(\bibinfo{publisher}{Cambridge University Press}, \bibinfo{year}{2001}).
	
	\bibitem[{\citenamefont{Van~Kampen}(1992)}]{van1992stochastic}
	\bibinfo{author}{\bibfnamefont{N.~G.} \bibnamefont{Van~Kampen}},
	\emph{\bibinfo{title}{Stochastic processes in physics and chemistry}},
	vol.~\bibinfo{volume}{1} (\bibinfo{publisher}{Elsevier},
	\bibinfo{year}{1992}).
	
	\bibitem[{\citenamefont{Klafter and Sokolov}(2011)}]{klafter2011first}
	\bibinfo{author}{\bibfnamefont{J.}~\bibnamefont{Klafter}} \bibnamefont{and}
	\bibinfo{author}{\bibfnamefont{I.~M.} \bibnamefont{Sokolov}},
	\emph{\bibinfo{title}{First steps in random walks: from tools to
			applications}} (\bibinfo{publisher}{Oxford University Press},
	\bibinfo{year}{2011}).
	
	\bibitem[{\citenamefont{Bray et~al.}(2013)\citenamefont{Bray, Majumdar, and
			Schehr}}]{bray2013persistence}
	\bibinfo{author}{\bibfnamefont{A.~J.} \bibnamefont{Bray}},
	\bibinfo{author}{\bibfnamefont{S.~N.} \bibnamefont{Majumdar}},
	\bibnamefont{and} \bibinfo{author}{\bibfnamefont{G.}~\bibnamefont{Schehr}},
	\bibinfo{journal}{Advances in Physics} \textbf{\bibinfo{volume}{62}},
	\bibinfo{pages}{225} (\bibinfo{year}{2013}).
	
	\bibitem[{\citenamefont{Bressloff and Newby}(2013)}]{RevModPhys.85.135}
	\bibinfo{author}{\bibfnamefont{P.~C.} \bibnamefont{Bressloff}}
	\bibnamefont{and} \bibinfo{author}{\bibfnamefont{J.~M.} \bibnamefont{Newby}},
	\bibinfo{journal}{Rev. Mod. Phys.} \textbf{\bibinfo{volume}{85}},
	\bibinfo{pages}{135} (\bibinfo{year}{2013}).
	
	\bibitem[{\citenamefont{Evans et~al.}(2020)\citenamefont{Evans, Majumdar, and
			Schehr}}]{evans2020stochastic}
	\bibinfo{author}{\bibfnamefont{M.~R.} \bibnamefont{Evans}},
	\bibinfo{author}{\bibfnamefont{S.~N.} \bibnamefont{Majumdar}},
	\bibnamefont{and} \bibinfo{author}{\bibfnamefont{G.}~\bibnamefont{Schehr}},
	\bibinfo{journal}{Journal of Physics A: Mathematical and Theoretical}
	\textbf{\bibinfo{volume}{53}}, \bibinfo{pages}{193001}
	(\bibinfo{year}{2020}).
	
	\bibitem[{\citenamefont{Kusmierz et~al.}(2014)\citenamefont{Kusmierz, Majumdar,
			Sabhapandit, and Schehr}}]{PhysRevLett.113.220602}
	\bibinfo{author}{\bibfnamefont{L.}~\bibnamefont{Kusmierz}},
	\bibinfo{author}{\bibfnamefont{S.~N.} \bibnamefont{Majumdar}},
	\bibinfo{author}{\bibfnamefont{S.}~\bibnamefont{Sabhapandit}},
	\bibnamefont{and} \bibinfo{author}{\bibfnamefont{G.}~\bibnamefont{Schehr}},
	\bibinfo{journal}{Phys. Rev. Lett.} \textbf{\bibinfo{volume}{113}},
	\bibinfo{pages}{220602} (\bibinfo{year}{2014}).
	
	\bibitem[{\citenamefont{Ku\ifmmode~\acute{s}\else \'{s}\fi{}mierz and
			Gudowska-Nowak}(2015)}]{PhysRevE.92.052127}
	\bibinfo{author}{\bibfnamefont{L.}~\bibnamefont{Ku\ifmmode~\acute{s}\else
			\'{s}\fi{}mierz}} \bibnamefont{and}
	\bibinfo{author}{\bibfnamefont{E.}~\bibnamefont{Gudowska-Nowak}},
	\bibinfo{journal}{Phys. Rev. E} \textbf{\bibinfo{volume}{92}},
	\bibinfo{pages}{052127} (\bibinfo{year}{2015}).
	
	\bibitem[{\citenamefont{Montanari and Zecchina}(2002)}]{PhysRevLett.88.178701}
	\bibinfo{author}{\bibfnamefont{A.}~\bibnamefont{Montanari}} \bibnamefont{and}
	\bibinfo{author}{\bibfnamefont{R.}~\bibnamefont{Zecchina}},
	\bibinfo{journal}{Phys. Rev. Lett.} \textbf{\bibinfo{volume}{88}},
	\bibinfo{pages}{178701} (\bibinfo{year}{2002}).
	
	\bibitem[{\citenamefont{Reuveni et~al.}(2014)\citenamefont{Reuveni, Urbakh, and
			Klafter}}]{reuveni2014role}
	\bibinfo{author}{\bibfnamefont{S.}~\bibnamefont{Reuveni}},
	\bibinfo{author}{\bibfnamefont{M.}~\bibnamefont{Urbakh}}, \bibnamefont{and}
	\bibinfo{author}{\bibfnamefont{J.}~\bibnamefont{Klafter}},
	\bibinfo{journal}{Proceedings of the National Academy of Sciences}
	\textbf{\bibinfo{volume}{111}}, \bibinfo{pages}{4391} (\bibinfo{year}{2014}).
	
	\bibitem[{\citenamefont{Rotbart et~al.}(2015)\citenamefont{Rotbart, Reuveni,
			and Urbakh}}]{rotbart2015michaelis}
	\bibinfo{author}{\bibfnamefont{T.}~\bibnamefont{Rotbart}},
	\bibinfo{author}{\bibfnamefont{S.}~\bibnamefont{Reuveni}}, \bibnamefont{and}
	\bibinfo{author}{\bibfnamefont{M.}~\bibnamefont{Urbakh}},
	\bibinfo{journal}{Physical Review E} \textbf{\bibinfo{volume}{92}},
	\bibinfo{pages}{060101} (\bibinfo{year}{2015}).
	
	\bibitem[{\citenamefont{Evans and
			Majumdar}(2011{\natexlab{a}})}]{evans2011diffusion}
	\bibinfo{author}{\bibfnamefont{M.~R.} \bibnamefont{Evans}} \bibnamefont{and}
	\bibinfo{author}{\bibfnamefont{S.~N.} \bibnamefont{Majumdar}},
	\bibinfo{journal}{Physical review letters} \textbf{\bibinfo{volume}{106}},
	\bibinfo{pages}{160601} (\bibinfo{year}{2011}{\natexlab{a}}).
	
	\bibitem[{\citenamefont{Evans and
			Majumdar}(2011{\natexlab{b}})}]{evans2011diffusion2}
	\bibinfo{author}{\bibfnamefont{M.~R.} \bibnamefont{Evans}} \bibnamefont{and}
	\bibinfo{author}{\bibfnamefont{S.~N.} \bibnamefont{Majumdar}},
	\bibinfo{journal}{Journal of Physics A: Mathematical and Theoretical}
	\textbf{\bibinfo{volume}{44}}, \bibinfo{pages}{435001}
	(\bibinfo{year}{2011}{\natexlab{b}}).
	
	\bibitem[{\citenamefont{Pal et~al.}(2016)\citenamefont{Pal, Kundu, and
			Evans}}]{pal2016diffusion}
	\bibinfo{author}{\bibfnamefont{A.}~\bibnamefont{Pal}},
	\bibinfo{author}{\bibfnamefont{A.}~\bibnamefont{Kundu}}, \bibnamefont{and}
	\bibinfo{author}{\bibfnamefont{M.~R.} \bibnamefont{Evans}},
	\bibinfo{journal}{Journal of Physics A: Mathematical and Theoretical}
	\textbf{\bibinfo{volume}{49}}, \bibinfo{pages}{225001}
	(\bibinfo{year}{2016}).
	
	\bibitem[{\citenamefont{Rold\'an and Gupta}(2017)}]{PhysRevE.96.022130}
	\bibinfo{author}{\bibfnamefont{E.}~\bibnamefont{Rold\'an}} \bibnamefont{and}
	\bibinfo{author}{\bibfnamefont{S.}~\bibnamefont{Gupta}},
	\bibinfo{journal}{Phys. Rev. E} \textbf{\bibinfo{volume}{96}},
	\bibinfo{pages}{022130} (\bibinfo{year}{2017}).
	
	\bibitem[{\citenamefont{Evans and
			Majumdar}(2014{\natexlab{a}})}]{Evans2014_Reset_Highd}
	\bibinfo{author}{\bibfnamefont{M.~R.} \bibnamefont{Evans}} \bibnamefont{and}
	\bibinfo{author}{\bibfnamefont{S.~N.} \bibnamefont{Majumdar}},
	\bibinfo{journal}{J. Phys. A: Math. Theor.} \textbf{\bibinfo{volume}{47}},
	\bibinfo{pages}{285001} (\bibinfo{year}{2014}{\natexlab{a}}).
	
	\bibitem[{\citenamefont{Chen and Huang}(2021)}]{arXiv:2109.11101}
	\bibinfo{author}{\bibfnamefont{H.}~\bibnamefont{Chen}} \bibnamefont{and}
	\bibinfo{author}{\bibfnamefont{F.}~\bibnamefont{Huang}}, p.
	\bibinfo{pages}{arXiv:2109.11101} (\bibinfo{year}{2021}).
	
	\bibitem[{\citenamefont{Christou and Schadschneider}(2015)}]{Christou2015}
	\bibinfo{author}{\bibfnamefont{C.}~\bibnamefont{Christou}} \bibnamefont{and}
	\bibinfo{author}{\bibfnamefont{A.}~\bibnamefont{Schadschneider}},
	\bibinfo{journal}{J. Phys. A: Math. Theor.} \textbf{\bibinfo{volume}{48}},
	\bibinfo{pages}{285003} (\bibinfo{year}{2015}).
	
	\bibitem[{\citenamefont{Domazetoski et~al.}(2020)\citenamefont{Domazetoski,
			Mas\'o-Puigdellosas, Sandev, M\'endez, Iomin, and
			Kocarev}}]{PhysRevResearch.2.033027}
	\bibinfo{author}{\bibfnamefont{V.}~\bibnamefont{Domazetoski}},
	\bibinfo{author}{\bibfnamefont{A.}~\bibnamefont{Mas\'o-Puigdellosas}},
	\bibinfo{author}{\bibfnamefont{T.}~\bibnamefont{Sandev}},
	\bibinfo{author}{\bibfnamefont{V.~m.~c.} \bibnamefont{M\'endez}},
	\bibinfo{author}{\bibfnamefont{A.}~\bibnamefont{Iomin}}, \bibnamefont{and}
	\bibinfo{author}{\bibfnamefont{L.}~\bibnamefont{Kocarev}},
	\bibinfo{journal}{Phys. Rev. Research} \textbf{\bibinfo{volume}{2}},
	\bibinfo{pages}{033027} (\bibinfo{year}{2020}).
	
	\bibitem[{\citenamefont{Bressloff}(2021)}]{BressloffJSTAT2021}
	\bibinfo{author}{\bibfnamefont{P.~C.} \bibnamefont{Bressloff}},
	\bibinfo{journal}{J. Stat. Mech.} p. \bibinfo{pages}{063206}
	(\bibinfo{year}{2021}).
	
	\bibitem[{\citenamefont{Evans and Majumdar}(2018{\natexlab{a}})}]{EvansJPA2018}
	\bibinfo{author}{\bibfnamefont{M.~R.} \bibnamefont{Evans}} \bibnamefont{and}
	\bibinfo{author}{\bibfnamefont{S.~N.} \bibnamefont{Majumdar}},
	\bibinfo{journal}{J. Phys. A: Math. Theor.} \textbf{\bibinfo{volume}{52}},
	\bibinfo{pages}{01LT01} (\bibinfo{year}{2018}{\natexlab{a}}).
	
	\bibitem[{\citenamefont{Pal et~al.}(2019)\citenamefont{Pal, Ku\'smierz, and
			Reuveni}}]{PalNJP2019}
	\bibinfo{author}{\bibfnamefont{A.}~\bibnamefont{Pal}},
	\bibinfo{author}{\bibfnamefont{L.}~\bibnamefont{Ku\'smierz}},
	\bibnamefont{and} \bibinfo{author}{\bibfnamefont{S.}~\bibnamefont{Reuveni}},
	\bibinfo{journal}{New J. Phys.} \textbf{\bibinfo{volume}{21}},
	\bibinfo{pages}{113024} (\bibinfo{year}{2019}).
	
	\bibitem[{\citenamefont{Bodrova and Sokolov}(2020)}]{PhysRevE.101.052130}
	\bibinfo{author}{\bibfnamefont{A.~S.} \bibnamefont{Bodrova}} \bibnamefont{and}
	\bibinfo{author}{\bibfnamefont{I.~M.} \bibnamefont{Sokolov}},
	\bibinfo{journal}{Phys. Rev. E} \textbf{\bibinfo{volume}{101}},
	\bibinfo{pages}{052130} (\bibinfo{year}{2020}).
	
	\bibitem[{\citenamefont{Gupta et~al.}(2020{\natexlab{a}})\citenamefont{Gupta,
			Plata, Kundu, and Pal}}]{GuptaJPA2020}
	\bibinfo{author}{\bibfnamefont{D.}~\bibnamefont{Gupta}},
	\bibinfo{author}{\bibfnamefont{C.~A.} \bibnamefont{Plata}},
	\bibinfo{author}{\bibfnamefont{A.}~\bibnamefont{Kundu}}, \bibnamefont{and}
	\bibinfo{author}{\bibfnamefont{A.}~\bibnamefont{Pal}}, \bibinfo{journal}{J.
		Phys. A: Math. Theor.} \textbf{\bibinfo{volume}{54}}, \bibinfo{pages}{025003}
	(\bibinfo{year}{2020}{\natexlab{a}}).
	
	\bibitem[{\citenamefont{Pal}(2015)}]{pal2015diffusion}
	\bibinfo{author}{\bibfnamefont{A.}~\bibnamefont{Pal}},
	\bibinfo{journal}{Physical Review E} \textbf{\bibinfo{volume}{91}},
	\bibinfo{pages}{012113} (\bibinfo{year}{2015}).
	
	\bibitem[{\citenamefont{Ahmad et~al.}(2019{\natexlab{a}})\citenamefont{Ahmad,
			Nayak, Bansal, Nandi, and Das}}]{ahmad2019first}
	\bibinfo{author}{\bibfnamefont{S.}~\bibnamefont{Ahmad}},
	\bibinfo{author}{\bibfnamefont{I.}~\bibnamefont{Nayak}},
	\bibinfo{author}{\bibfnamefont{A.}~\bibnamefont{Bansal}},
	\bibinfo{author}{\bibfnamefont{A.}~\bibnamefont{Nandi}}, \bibnamefont{and}
	\bibinfo{author}{\bibfnamefont{D.}~\bibnamefont{Das}},
	\bibinfo{journal}{Physical Review E} \textbf{\bibinfo{volume}{99}},
	\bibinfo{pages}{022130} (\bibinfo{year}{2019}{\natexlab{a}}).
	
	\bibitem[{\citenamefont{Gupta et~al.}(2020{\natexlab{b}})\citenamefont{Gupta,
			Plata, Kundu, and Pal}}]{gupta2020stochastic}
	\bibinfo{author}{\bibfnamefont{D.}~\bibnamefont{Gupta}},
	\bibinfo{author}{\bibfnamefont{C.~A.} \bibnamefont{Plata}},
	\bibinfo{author}{\bibfnamefont{A.}~\bibnamefont{Kundu}}, \bibnamefont{and}
	\bibinfo{author}{\bibfnamefont{A.}~\bibnamefont{Pal}},
	\bibinfo{journal}{Journal of Physics A: Mathematical and Theoretical}
	\textbf{\bibinfo{volume}{54}}, \bibinfo{pages}{025003}
	(\bibinfo{year}{2020}{\natexlab{b}}).
	
	\bibitem[{\citenamefont{Pal and
			Prasad}(2019{\natexlab{a}})}]{PhysRevE.99.032123}
	\bibinfo{author}{\bibfnamefont{A.}~\bibnamefont{Pal}} \bibnamefont{and}
	\bibinfo{author}{\bibfnamefont{V.~V.} \bibnamefont{Prasad}},
	\bibinfo{journal}{Phys. Rev. E} \textbf{\bibinfo{volume}{99}},
	\bibinfo{pages}{032123} (\bibinfo{year}{2019}{\natexlab{a}}).
	
	\bibitem[{\citenamefont{Bressloff}(2020{\natexlab{a}})}]{Bressloff2020JPA1}
	\bibinfo{author}{\bibfnamefont{P.~C.} \bibnamefont{Bressloff}},
	\bibinfo{journal}{J. Phys. A: Math. Theor.} \textbf{\bibinfo{volume}{53}},
	\bibinfo{pages}{105001} (\bibinfo{year}{2020}{\natexlab{a}}).
	
	\bibitem[{\citenamefont{Bressloff}(2020{\natexlab{b}})}]{PhysRevE.102.022115}
	\bibinfo{author}{\bibfnamefont{P.~C.} \bibnamefont{Bressloff}},
	\bibinfo{journal}{Phys. Rev. E} \textbf{\bibinfo{volume}{102}},
	\bibinfo{pages}{022115} (\bibinfo{year}{2020}{\natexlab{b}}).
	
	\bibitem[{\citenamefont{Evans and Majumdar}(2018{\natexlab{b}})}]{evans2018run}
	\bibinfo{author}{\bibfnamefont{M.~R.} \bibnamefont{Evans}} \bibnamefont{and}
	\bibinfo{author}{\bibfnamefont{S.~N.} \bibnamefont{Majumdar}},
	\bibinfo{journal}{Journal of Physics A: Mathematical and Theoretical}
	\textbf{\bibinfo{volume}{51}}, \bibinfo{pages}{475003}
	(\bibinfo{year}{2018}{\natexlab{b}}).
	
	\bibitem[{\citenamefont{Santra et~al.}(2020)\citenamefont{Santra, Basu, and
			Sabhapandit}}]{santra2020run}
	\bibinfo{author}{\bibfnamefont{I.}~\bibnamefont{Santra}},
	\bibinfo{author}{\bibfnamefont{U.}~\bibnamefont{Basu}}, \bibnamefont{and}
	\bibinfo{author}{\bibfnamefont{S.}~\bibnamefont{Sabhapandit}},
	\bibinfo{journal}{Journal of Statistical Mechanics: Theory and Experiment}
	\textbf{\bibinfo{volume}{2020}}, \bibinfo{pages}{113206}
	(\bibinfo{year}{2020}).
	
	\bibitem[{\citenamefont{Bressloff}(2020{\natexlab{c}})}]{bressloff2020occupation}
	\bibinfo{author}{\bibfnamefont{P.~C.} \bibnamefont{Bressloff}},
	\bibinfo{journal}{Physical Review E} \textbf{\bibinfo{volume}{102}},
	\bibinfo{pages}{042135} (\bibinfo{year}{2020}{\natexlab{c}}).
	
	\bibitem[{\citenamefont{Scacchi and Sharma}(2018)}]{scacchi2018mean}
	\bibinfo{author}{\bibfnamefont{A.}~\bibnamefont{Scacchi}} \bibnamefont{and}
	\bibinfo{author}{\bibfnamefont{A.}~\bibnamefont{Sharma}},
	\bibinfo{journal}{Molecular Physics} \textbf{\bibinfo{volume}{116}},
	\bibinfo{pages}{460} (\bibinfo{year}{2018}).
	
	\bibitem[{\citenamefont{Kumar et~al.}(2020)\citenamefont{Kumar, Sadekar, and
			Basu}}]{kumar2020active}
	\bibinfo{author}{\bibfnamefont{V.}~\bibnamefont{Kumar}},
	\bibinfo{author}{\bibfnamefont{O.}~\bibnamefont{Sadekar}}, \bibnamefont{and}
	\bibinfo{author}{\bibfnamefont{U.}~\bibnamefont{Basu}},
	\bibinfo{journal}{Physical Review E} \textbf{\bibinfo{volume}{102}},
	\bibinfo{pages}{052129} (\bibinfo{year}{2020}).
	
	\bibitem[{\citenamefont{Basu et~al.}(2019)\citenamefont{Basu, Kundu, and
			Pal}}]{basu2019symmetric}
	\bibinfo{author}{\bibfnamefont{U.}~\bibnamefont{Basu}},
	\bibinfo{author}{\bibfnamefont{A.}~\bibnamefont{Kundu}}, \bibnamefont{and}
	\bibinfo{author}{\bibfnamefont{A.}~\bibnamefont{Pal}},
	\bibinfo{journal}{Physical Review E} \textbf{\bibinfo{volume}{100}},
	\bibinfo{pages}{032136} (\bibinfo{year}{2019}).
	
	\bibitem[{\citenamefont{Pal and Reuveni}(2017)}]{pal2017first}
	\bibinfo{author}{\bibfnamefont{A.}~\bibnamefont{Pal}} \bibnamefont{and}
	\bibinfo{author}{\bibfnamefont{S.}~\bibnamefont{Reuveni}},
	\bibinfo{journal}{Physical review letters} \textbf{\bibinfo{volume}{118}},
	\bibinfo{pages}{030603} (\bibinfo{year}{2017}).
	
	\bibitem[{\citenamefont{Gupta et~al.}(2014)\citenamefont{Gupta, Majumdar, and
			Schehr}}]{gupta2014fluctuating}
	\bibinfo{author}{\bibfnamefont{S.}~\bibnamefont{Gupta}},
	\bibinfo{author}{\bibfnamefont{S.~N.} \bibnamefont{Majumdar}},
	\bibnamefont{and} \bibinfo{author}{\bibfnamefont{G.}~\bibnamefont{Schehr}},
	\bibinfo{journal}{Physical review letters} \textbf{\bibinfo{volume}{112}},
	\bibinfo{pages}{220601} (\bibinfo{year}{2014}).
	
	\bibitem[{\citenamefont{Evans and
			Majumdar}(2014{\natexlab{b}})}]{evans2014diffusion}
	\bibinfo{author}{\bibfnamefont{M.~R.} \bibnamefont{Evans}} \bibnamefont{and}
	\bibinfo{author}{\bibfnamefont{S.~N.} \bibnamefont{Majumdar}},
	\bibinfo{journal}{Journal of Physics A: Mathematical and Theoretical}
	\textbf{\bibinfo{volume}{47}}, \bibinfo{pages}{285001}
	(\bibinfo{year}{2014}{\natexlab{b}}).
	
	\bibitem[{\citenamefont{Meylahn et~al.}(2015)\citenamefont{Meylahn,
			Sabhapandit, and Touchette}}]{meylahn2015large}
	\bibinfo{author}{\bibfnamefont{J.~M.} \bibnamefont{Meylahn}},
	\bibinfo{author}{\bibfnamefont{S.}~\bibnamefont{Sabhapandit}},
	\bibnamefont{and}
	\bibinfo{author}{\bibfnamefont{H.}~\bibnamefont{Touchette}},
	\bibinfo{journal}{Physical Review E} \textbf{\bibinfo{volume}{92}},
	\bibinfo{pages}{062148} (\bibinfo{year}{2015}).
	
	\bibitem[{\citenamefont{Chechkin and Sokolov}(2018)}]{chechkin2018random}
	\bibinfo{author}{\bibfnamefont{A.}~\bibnamefont{Chechkin}} \bibnamefont{and}
	\bibinfo{author}{\bibfnamefont{I.}~\bibnamefont{Sokolov}},
	\bibinfo{journal}{Physical review letters} \textbf{\bibinfo{volume}{121}},
	\bibinfo{pages}{050601} (\bibinfo{year}{2018}).
	
	\bibitem[{\citenamefont{Magoni et~al.}(2020)\citenamefont{Magoni, Majumdar, and
			Schehr}}]{magoni2020ising}
	\bibinfo{author}{\bibfnamefont{M.}~\bibnamefont{Magoni}},
	\bibinfo{author}{\bibfnamefont{S.~N.} \bibnamefont{Majumdar}},
	\bibnamefont{and} \bibinfo{author}{\bibfnamefont{G.}~\bibnamefont{Schehr}},
	\bibinfo{journal}{Physical Review Research} \textbf{\bibinfo{volume}{2}},
	\bibinfo{pages}{033182} (\bibinfo{year}{2020}).
	
	\bibitem[{\citenamefont{Fuchs et~al.}(2016)\citenamefont{Fuchs, Goldt, and
			Seifert}}]{fuchs2016stochastic}
	\bibinfo{author}{\bibfnamefont{J.}~\bibnamefont{Fuchs}},
	\bibinfo{author}{\bibfnamefont{S.}~\bibnamefont{Goldt}}, \bibnamefont{and}
	\bibinfo{author}{\bibfnamefont{U.}~\bibnamefont{Seifert}},
	\bibinfo{journal}{EPL (Europhysics Letters)} \textbf{\bibinfo{volume}{113}},
	\bibinfo{pages}{60009} (\bibinfo{year}{2016}).
	
	\bibitem[{\citenamefont{Pal and Rahav}(2017)}]{pal2017integral}
	\bibinfo{author}{\bibfnamefont{A.}~\bibnamefont{Pal}} \bibnamefont{and}
	\bibinfo{author}{\bibfnamefont{S.}~\bibnamefont{Rahav}},
	\bibinfo{journal}{Physical Review E} \textbf{\bibinfo{volume}{96}},
	\bibinfo{pages}{062135} (\bibinfo{year}{2017}).
	
	\bibitem[{\citenamefont{Gupta et~al.}(2020{\natexlab{c}})\citenamefont{Gupta,
			Plata, and Pal}}]{gupta2020work}
	\bibinfo{author}{\bibfnamefont{D.}~\bibnamefont{Gupta}},
	\bibinfo{author}{\bibfnamefont{C.~A.} \bibnamefont{Plata}}, \bibnamefont{and}
	\bibinfo{author}{\bibfnamefont{A.}~\bibnamefont{Pal}},
	\bibinfo{journal}{Physical review letters} \textbf{\bibinfo{volume}{124}},
	\bibinfo{pages}{110608} (\bibinfo{year}{2020}{\natexlab{c}}).
	
	\bibitem[{\citenamefont{Tal-Friedman et~al.}(2020)\citenamefont{Tal-Friedman,
			Pal, Sekhon, Reuveni, and Roichman}}]{tal2020experimental}
	\bibinfo{author}{\bibfnamefont{O.}~\bibnamefont{Tal-Friedman}},
	\bibinfo{author}{\bibfnamefont{A.}~\bibnamefont{Pal}},
	\bibinfo{author}{\bibfnamefont{A.}~\bibnamefont{Sekhon}},
	\bibinfo{author}{\bibfnamefont{S.}~\bibnamefont{Reuveni}}, \bibnamefont{and}
	\bibinfo{author}{\bibfnamefont{Y.}~\bibnamefont{Roichman}},
	\bibinfo{journal}{The journal of physical chemistry letters}
	\textbf{\bibinfo{volume}{11}}, \bibinfo{pages}{7350} (\bibinfo{year}{2020}).
	
	\bibitem[{\citenamefont{Besga et~al.}(2020)\citenamefont{Besga, Bovon,
			Petrosyan, Majumdar, and Ciliberto}}]{besga2020optimal}
	\bibinfo{author}{\bibfnamefont{B.}~\bibnamefont{Besga}},
	\bibinfo{author}{\bibfnamefont{A.}~\bibnamefont{Bovon}},
	\bibinfo{author}{\bibfnamefont{A.}~\bibnamefont{Petrosyan}},
	\bibinfo{author}{\bibfnamefont{S.~N.} \bibnamefont{Majumdar}},
	\bibnamefont{and}
	\bibinfo{author}{\bibfnamefont{S.}~\bibnamefont{Ciliberto}},
	\bibinfo{journal}{Physical Review Research} \textbf{\bibinfo{volume}{2}},
	\bibinfo{pages}{032029} (\bibinfo{year}{2020}).
	
	\bibitem[{\citenamefont{Reuveni}(2016)}]{PhysRevLett.116.170601}
	\bibinfo{author}{\bibfnamefont{S.}~\bibnamefont{Reuveni}},
	\bibinfo{journal}{Phys. Rev. Lett.} \textbf{\bibinfo{volume}{116}},
	\bibinfo{pages}{170601} (\bibinfo{year}{2016}).
	
	\bibitem[{\citenamefont{Pal et~al.}(2021)\citenamefont{Pal, Kostinski, and
			Reuveni}}]{Pal2021}
	\bibinfo{author}{\bibfnamefont{A.}~\bibnamefont{Pal}},
	\bibinfo{author}{\bibfnamefont{S.}~\bibnamefont{Kostinski}},
	\bibnamefont{and} \bibinfo{author}{\bibfnamefont{S.}~\bibnamefont{Reuveni}},
	\bibinfo{journal}{arXiv:2108.07018}  (\bibinfo{year}{2021}).
	
	\bibitem[{\citenamefont{Ray et~al.}(2019)\citenamefont{Ray, Mondal, and
			Reuveni}}]{Ray2019}
	\bibinfo{author}{\bibfnamefont{S.}~\bibnamefont{Ray}},
	\bibinfo{author}{\bibfnamefont{D.}~\bibnamefont{Mondal}}, \bibnamefont{and}
	\bibinfo{author}{\bibfnamefont{S.}~\bibnamefont{Reuveni}},
	\bibinfo{journal}{J. Phys. A: Math. Theor.} \textbf{\bibinfo{volume}{52}},
	\bibinfo{pages}{255002} (\bibinfo{year}{2019}).
	
	\bibitem[{\citenamefont{Durang et~al.}(2019)\citenamefont{Durang, Lee, Lizana,
			and Jeon}}]{Durang2019}
	\bibinfo{author}{\bibfnamefont{X.}~\bibnamefont{Durang}},
	\bibinfo{author}{\bibfnamefont{S.}~\bibnamefont{Lee}},
	\bibinfo{author}{\bibfnamefont{L.}~\bibnamefont{Lizana}}, \bibnamefont{and}
	\bibinfo{author}{\bibfnamefont{J.-H.} \bibnamefont{Jeon}},
	\bibinfo{journal}{J. Phys. A: Math. Theor.} \textbf{\bibinfo{volume}{52}},
	\bibinfo{pages}{224001} (\bibinfo{year}{2019}).
	
	\bibitem[{\citenamefont{Pal and
			Prasad}(2019{\natexlab{b}})}]{PhysRevResearch.1.032001}
	\bibinfo{author}{\bibfnamefont{A.}~\bibnamefont{Pal}} \bibnamefont{and}
	\bibinfo{author}{\bibfnamefont{V.~V.} \bibnamefont{Prasad}},
	\bibinfo{journal}{Phys. Rev. Research} \textbf{\bibinfo{volume}{1}},
	\bibinfo{pages}{032001} (\bibinfo{year}{2019}{\natexlab{b}}).
	
	\bibitem[{\citenamefont{Ahmad et~al.}(2019{\natexlab{b}})\citenamefont{Ahmad,
			Nayak, Bansal, Nandi, and Das}}]{PhysRevE.99.022130}
	\bibinfo{author}{\bibfnamefont{S.}~\bibnamefont{Ahmad}},
	\bibinfo{author}{\bibfnamefont{I.}~\bibnamefont{Nayak}},
	\bibinfo{author}{\bibfnamefont{A.}~\bibnamefont{Bansal}},
	\bibinfo{author}{\bibfnamefont{A.}~\bibnamefont{Nandi}}, \bibnamefont{and}
	\bibinfo{author}{\bibfnamefont{D.}~\bibnamefont{Das}},
	\bibinfo{journal}{Phys. Rev. E} \textbf{\bibinfo{volume}{99}},
	\bibinfo{pages}{022130} (\bibinfo{year}{2019}{\natexlab{b}}).
	
	\bibitem[{\citenamefont{Bonomo and Pal}(2021)}]{PhysRevE.103.052129}
	\bibinfo{author}{\bibfnamefont{O.~L.} \bibnamefont{Bonomo}} \bibnamefont{and}
	\bibinfo{author}{\bibfnamefont{A.}~\bibnamefont{Pal}},
	\bibinfo{journal}{Phys. Rev. E} \textbf{\bibinfo{volume}{103}},
	\bibinfo{pages}{052129} (\bibinfo{year}{2021}).
	
	\bibitem[{\citenamefont{Campos and M\'endez}(2015)}]{PhysRevE.92.062115}
	\bibinfo{author}{\bibfnamefont{D.}~\bibnamefont{Campos}} \bibnamefont{and}
	\bibinfo{author}{\bibfnamefont{V.~m.~c.} \bibnamefont{M\'endez}},
	\bibinfo{journal}{Phys. Rev. E} \textbf{\bibinfo{volume}{92}},
	\bibinfo{pages}{062115} (\bibinfo{year}{2015}).
	
	\bibitem[{\citenamefont{Chatterjee et~al.}(2018)\citenamefont{Chatterjee,
			Christou, and Schadschneider}}]{PhysRevE.97.062106}
	\bibinfo{author}{\bibfnamefont{A.}~\bibnamefont{Chatterjee}},
	\bibinfo{author}{\bibfnamefont{C.}~\bibnamefont{Christou}}, \bibnamefont{and}
	\bibinfo{author}{\bibfnamefont{A.}~\bibnamefont{Schadschneider}},
	\bibinfo{journal}{Phys. Rev. E} \textbf{\bibinfo{volume}{97}},
	\bibinfo{pages}{062106} (\bibinfo{year}{2018}).
	
	\bibitem[{\citenamefont{Montero and Villarroel}(2016)}]{PhysRevE.94.032132}
	\bibinfo{author}{\bibfnamefont{M.}~\bibnamefont{Montero}} \bibnamefont{and}
	\bibinfo{author}{\bibfnamefont{J.}~\bibnamefont{Villarroel}},
	\bibinfo{journal}{Phys. Rev. E} \textbf{\bibinfo{volume}{94}},
	\bibinfo{pages}{032132} (\bibinfo{year}{2016}).
	
	\bibitem[{\citenamefont{Boyer and Solis-Salas}(2014)}]{PhysRevLett.112.240601}
	\bibinfo{author}{\bibfnamefont{D.}~\bibnamefont{Boyer}} \bibnamefont{and}
	\bibinfo{author}{\bibfnamefont{C.}~\bibnamefont{Solis-Salas}},
	\bibinfo{journal}{Phys. Rev. Lett.} \textbf{\bibinfo{volume}{112}},
	\bibinfo{pages}{240601} (\bibinfo{year}{2014}).
	
	\bibitem[{\citenamefont{Falc\'on-Cort\'es
			et~al.}(2017)\citenamefont{Falc\'on-Cort\'es, Boyer, Giuggioli, and
			Majumdar}}]{PhysRevLett.119.140603}
	\bibinfo{author}{\bibfnamefont{A.}~\bibnamefont{Falc\'on-Cort\'es}},
	\bibinfo{author}{\bibfnamefont{D.}~\bibnamefont{Boyer}},
	\bibinfo{author}{\bibfnamefont{L.}~\bibnamefont{Giuggioli}},
	\bibnamefont{and} \bibinfo{author}{\bibfnamefont{S.~N.}
		\bibnamefont{Majumdar}}, \bibinfo{journal}{Phys. Rev. Lett.}
	\textbf{\bibinfo{volume}{119}}, \bibinfo{pages}{140603}
	(\bibinfo{year}{2017}).
	
	\bibitem[{\citenamefont{Boyer et~al.}(2019)\citenamefont{Boyer,
			Falc\'n-Cort\'es, Giuggioli, and Majumdar}}]{Boyer2019}
	\bibinfo{author}{\bibfnamefont{D.}~\bibnamefont{Boyer}},
	\bibinfo{author}{\bibfnamefont{A.}~\bibnamefont{Falc\'n-Cort\'es}},
	\bibinfo{author}{\bibfnamefont{L.}~\bibnamefont{Giuggioli}},
	\bibnamefont{and} \bibinfo{author}{\bibfnamefont{S.~N.}
		\bibnamefont{Majumdar}}, \bibinfo{journal}{J. Stat. Mech.}
	\textbf{\bibinfo{volume}{2019}}, \bibinfo{pages}{053204}
	(\bibinfo{year}{2019}).
	
	\bibitem[{\citenamefont{Majumdar et~al.}(2015)\citenamefont{Majumdar,
			Sabhapandit, and Schehr}}]{PhysRevE.92.052126}
	\bibinfo{author}{\bibfnamefont{S.~N.} \bibnamefont{Majumdar}},
	\bibinfo{author}{\bibfnamefont{S.}~\bibnamefont{Sabhapandit}},
	\bibnamefont{and} \bibinfo{author}{\bibfnamefont{G.}~\bibnamefont{Schehr}},
	\bibinfo{journal}{Phys. Rev. E} \textbf{\bibinfo{volume}{92}},
	\bibinfo{pages}{052126} (\bibinfo{year}{2015}).
	
	\bibitem[{\citenamefont{Riascos et~al.}(2020)\citenamefont{Riascos, Boyer,
			Herringer, and Mateos}}]{PhysRevE.101.062147}
	\bibinfo{author}{\bibfnamefont{A.~P.} \bibnamefont{Riascos}},
	\bibinfo{author}{\bibfnamefont{D.}~\bibnamefont{Boyer}},
	\bibinfo{author}{\bibfnamefont{P.}~\bibnamefont{Herringer}},
	\bibnamefont{and} \bibinfo{author}{\bibfnamefont{J.~L.}
		\bibnamefont{Mateos}}, \bibinfo{journal}{Phys. Rev. E}
	\textbf{\bibinfo{volume}{101}}, \bibinfo{pages}{062147}
	(\bibinfo{year}{2020}).
	
	\bibitem[{\citenamefont{Gonz\'alez et~al.}(2021)\citenamefont{Gonz\'alez,
			Riascos, and Boyer}}]{PhysRevE.103.062126}
	\bibinfo{author}{\bibfnamefont{F.~H.} \bibnamefont{Gonz\'alez}},
	\bibinfo{author}{\bibfnamefont{A.~P.} \bibnamefont{Riascos}},
	\bibnamefont{and} \bibinfo{author}{\bibfnamefont{D.}~\bibnamefont{Boyer}},
	\bibinfo{journal}{Phys. Rev. E} \textbf{\bibinfo{volume}{103}},
	\bibinfo{pages}{062126} (\bibinfo{year}{2021}).
	
	\bibitem[{\citenamefont{Wang et~al.}(2021)\citenamefont{Wang, Chen, and
			Huang}}]{Wang2021}
	\bibinfo{author}{\bibfnamefont{S.}~\bibnamefont{Wang}},
	\bibinfo{author}{\bibfnamefont{H.}~\bibnamefont{Chen}}, \bibnamefont{and}
	\bibinfo{author}{\bibfnamefont{F.}~\bibnamefont{Huang}},
	\bibinfo{journal}{Chaos} \textbf{\bibinfo{volume}{31}},
	\bibinfo{pages}{093135} (\bibinfo{year}{2021}).
	
	\bibitem[{\citenamefont{Wald and B\"ottcher}(2021)}]{PhysRevE.103.012122}
	\bibinfo{author}{\bibfnamefont{S.}~\bibnamefont{Wald}} \bibnamefont{and}
	\bibinfo{author}{\bibfnamefont{L.}~\bibnamefont{B\"ottcher}},
	\bibinfo{journal}{Phys. Rev. E} \textbf{\bibinfo{volume}{103}},
	\bibinfo{pages}{012122} (\bibinfo{year}{2021}).
	
	\bibitem[{\citenamefont{Huang and Chen}(2021)}]{huang2021random}
	\bibinfo{author}{\bibfnamefont{F.}~\bibnamefont{Huang}} \bibnamefont{and}
	\bibinfo{author}{\bibfnamefont{H.}~\bibnamefont{Chen}},
	\bibinfo{journal}{Phys. Rev. E} \textbf{\bibinfo{volume}{103}},
	\bibinfo{pages}{062132} (\bibinfo{year}{2021}).
	
	\bibitem[{\citenamefont{De~Bruyne et~al.}(2020)\citenamefont{De~Bruyne,
			Randon-Furling, and Redner}}]{de2020optimization}
	\bibinfo{author}{\bibfnamefont{B.}~\bibnamefont{De~Bruyne}},
	\bibinfo{author}{\bibfnamefont{J.}~\bibnamefont{Randon-Furling}},
	\bibnamefont{and} \bibinfo{author}{\bibfnamefont{S.}~\bibnamefont{Redner}},
	\bibinfo{journal}{Physical Review Letters} \textbf{\bibinfo{volume}{125}},
	\bibinfo{pages}{050602} (\bibinfo{year}{2020}).
	
	\bibitem[{\citenamefont{De~Bruyne et~al.}(2021)\citenamefont{De~Bruyne,
			Randon-Furling, and Redner}}]{de2021optimization}
	\bibinfo{author}{\bibfnamefont{B.}~\bibnamefont{De~Bruyne}},
	\bibinfo{author}{\bibfnamefont{J.}~\bibnamefont{Randon-Furling}},
	\bibnamefont{and} \bibinfo{author}{\bibfnamefont{S.}~\bibnamefont{Redner}},
	\bibinfo{journal}{Journal of Statistical Mechanics: Theory and Experiment}
	\textbf{\bibinfo{volume}{2021}}, \bibinfo{pages}{013203}
	(\bibinfo{year}{2021}).
	
	\bibitem[{\citenamefont{Usmani}(1994)}]{usmani1994inversion}
	\bibinfo{author}{\bibfnamefont{R.~A.} \bibnamefont{Usmani}},
	\bibinfo{journal}{Linear Algebra and its Applications}
	\textbf{\bibinfo{volume}{212}}, \bibinfo{pages}{413} (\bibinfo{year}{1994}).
	
	\bibitem[{\citenamefont{Sood and Redner}(2005)}]{PhysRevLett.94.178701}
	\bibinfo{author}{\bibfnamefont{V.}~\bibnamefont{Sood}} \bibnamefont{and}
	\bibinfo{author}{\bibfnamefont{S.}~\bibnamefont{Redner}},
	\bibinfo{journal}{Phys. Rev. Lett.} \textbf{\bibinfo{volume}{94}},
	\bibinfo{pages}{178701} (\bibinfo{year}{2005}).
	
	\bibitem[{\citenamefont{Sood et~al.}(2008)\citenamefont{Sood, Antal, and
			Redner}}]{PhysRevE.77.041121}
	\bibinfo{author}{\bibfnamefont{V.}~\bibnamefont{Sood}},
	\bibinfo{author}{\bibfnamefont{T.}~\bibnamefont{Antal}}, \bibnamefont{and}
	\bibinfo{author}{\bibfnamefont{S.}~\bibnamefont{Redner}},
	\bibinfo{journal}{Phys. Rev. E} \textbf{\bibinfo{volume}{77}},
	\bibinfo{pages}{041121} (\bibinfo{year}{2008}).
	
	\bibitem[{\citenamefont{ben Avraham et~al.}(1990)\citenamefont{ben Avraham,
			Considine, Meakin, Redner, and Takayasu}}]{AvrahamJPA1990}
	\bibinfo{author}{\bibfnamefont{D.}~\bibnamefont{ben Avraham}},
	\bibinfo{author}{\bibfnamefont{D.}~\bibnamefont{Considine}},
	\bibinfo{author}{\bibfnamefont{P.}~\bibnamefont{Meakin}},
	\bibinfo{author}{\bibfnamefont{S.}~\bibnamefont{Redner}}, \bibnamefont{and}
	\bibinfo{author}{\bibfnamefont{H.}~\bibnamefont{Takayasu}},
	\bibinfo{journal}{J. Phys. A: Math. Theor.} \textbf{\bibinfo{volume}{23}},
	\bibinfo{pages}{4297} (\bibinfo{year}{1990}).
	
	\bibitem[{\citenamefont{Slanina and Lavicka}(2003)}]{SlaninaEPJB2003}
	\bibinfo{author}{\bibfnamefont{F.}~\bibnamefont{Slanina}} \bibnamefont{and}
	\bibinfo{author}{\bibfnamefont{H.}~\bibnamefont{Lavicka}},
	\bibinfo{journal}{Eur. Phys. J. B} \textbf{\bibinfo{volume}{35}},
	\bibinfo{pages}{279} (\bibinfo{year}{2003}).
	
	\bibitem[{\citenamefont{Vazquez and Egu\'iluz}(2008)}]{VazquezNJP2008}
	\bibinfo{author}{\bibfnamefont{F.}~\bibnamefont{Vazquez}} \bibnamefont{and}
	\bibinfo{author}{\bibfnamefont{V.~M.} \bibnamefont{Egu\'iluz}},
	\bibinfo{journal}{New J. Phys.} \textbf{\bibinfo{volume}{10}},
	\bibinfo{pages}{063011} (\bibinfo{year}{2008}).
	
\end{thebibliography}
\end{document}